%% file: XorderedmPDFs.tex
\documentclass[a4paper,11pt]{article}
\usepackage{jheppub} 
\usepackage[T1]{fontenc} 
\usepackage{graphicx}
\usepackage{bm}
\usepackage{amsmath}
\usepackage{amssymb}
\usepackage{url}
\usepackage{epsfig}
\usepackage[T1]{fontenc}
\usepackage[latin9]{inputenc}
\usepackage{multirow}
\usepackage{xspace}
\usepackage{xcolor}
\usepackage{mathtools}
\usepackage[justification=raggedright]{caption}
\usepackage{hyperref}
\usepackage{float}
\usepackage{microtype}
\usepackage{cprotect}
\usepackage{makecell}
\usepackage{float}
\usepackage{subfigure}

\usepackage{tabu}

\usepackage{listings}
\input{preamble}

\newcommand*\phantomrel[1]{\mathrel{\phantom{#1}}}

\def\beq{\begin{equation}}
\def\eeq{\end{equation}}
\def\bear{\begin{eqnarray}}
\def\eear{\end{eqnarray}}

\title{Improving on the \pythia modelling of equal-scale multi-parton distribution functions}

\author[a, b, c]{Oleh Fedkevych,} 
\author[d, e]{Jonathan R. Gaunt,}
\author[d]{Seonagh Smith}

\affiliation[a]{Physics and Astronomy Department, Georgia State University, Atlanta, GA 30303, USA}

\affiliation[b]{University of Jyvaskyla, Department of Physics, P.O. Box 35, FI-40014 University of Jyvaskyla, Finland}

\affiliation[c]{Helsinki Institute of Physics, P.O. Box 64, FI-00014 University of Helsinki, Finland}

\affiliation[d]{Department of Physics and Astronomy, University of Manchester, Manchester, M13 9PL, United Kingdom}
\affiliation[e]{Department of Physics, University of Cyprus, Nicosia 1678, Cyprus}

\emailAdd{oleh.o.fedkevych@jyu.fi}
\emailAdd{gaunt.jonathan@ucy.ac.cy}

\abstract{Multi-parton distribution functions (mPDFs) are non-perturbative objects that are important in the prediction of multiple scattering rates at hadron colliders. In the case where the scales associated with all partons in the mPDF are the same, we have two theoretical constraints on the mPDF. These are symmetry in exchange of the parton indices, and the number and momentum sum rules. In a previous publication \cite{Fedkevych:2022myf} we found that the equal-scale mPDFs from the \pythia model could not satisfy both of these constraints simultaneously. In this paper we introduce an algorithm for constructing equal-scale mPDFs that is based on the \pythia procedure but has three additional modifications, such that it yields symmetric mPDFs that should satisfy the sum rules to an improved extent. We test the construction for the case of the double and triple parton distribution functions (dPDFs and tPDFs), finding that the sum rules are obeyed to within $10\%$ over the vast majority of the phase space for the scales tested (and deviations only being mildly above this level). We use our dPDFs to compute rapidity asymmetries for same-sign $WW$ and $ZZ$ production via double parton scattering, and compare the results to predictions obtained using \pythia and the GS09 dPDFs of Ref.~\cite{Gaunt:2009re}.}

\begin{document}

\vspace{1cm}

\maketitle
\setcounter{footnote}{0}

\section{Introduction}
\label{s:intro}

Multiple parton scattering (MPI) is the process where multiple parton-parton scatterings occur in an individual proton-proton collision. Multiple low-energy scatters accompanying a trigger process are typical at hadron colliders such as the LHC and form the underlying event, which impacts a variety of measurements (such as jet measurements). Multiple hard parton-parton scatters are rarer, but can compete with single scattering signals for certain processes and/or phase space regions (see \textit{e.g.}~Refs.~\cite{Kulesza:1999zh,Gaunt:2010pi, Kom:2011bd,Diehl:2011tt, Diehl:2011yj, Helenius:2019uge, Fedkevych:2020cmd}). Thus, theoretical predictions of multiple scattering are relevant to a variety of measurements at the LHC (see \textit{e.g.} ~Ref.~\cite{Bartalini:2018qje} for a summary). Multiple interactions are interesting in their own right as they reveal novel information on hadron structure, in particular the correlations between partons in a hadron. Various types of possible parton correlations have been studied (see \textit{e.g.}~Refs.~\cite{Korotkikh:2004bz,Cattaruzza:2005nu,Gaunt:2009re,Calucci:1997ii,Calucci:2010wg,Rogers:2009ke, Domdey:2009bg,Flensburg:2011kj,Blok:2012mw,Kasemets:2012pr,Blok:2013bpa, Diehl:2017kgu, Cotogno:2018mfv, Cotogno:2020iio, Huayra:2023gio,Lovato:2025jgh, Dumitru:2025bib}).

An important object in the theoretical description of multiple scattering is the \textit{$m$-parton distribution function}, or mPDF. Roughly speaking, this object expresses the probability to find $m$ partons in the proton with given momentum fractions $x_i$ and flavours $j_i$ at given scales $Q_i$  ($i\in\{1,m\}$), encoding ``longitudinal'' correlations in momentum fraction between the partons. This object does not directly appear in proton-proton multiple scattering cross sections -- the one that appears there is the multiple parton distribution (MPD) that also encodes transverse correlations via dependence on the parton transverse separations  $\bm{y}_i$ \cite{Diehl:2011tt, Diehl:2017kgu}. However, the two are related by an integral over transverse separations, and since the mPDF is a simpler object we can regard it as an important ``stepping stone'' on the way to the description of the full MPD. For simplicity, let us restrict our attention here to the case of the equal-scale mPDF, relevant to the case in which all $Q_i$ are equal, $Q_i = Q$.

Even the equal-scale mPDFs are complex nonperturbative objects that are a function of several arguments, and either computing them from first principles or fitting them from data remains out of reach for the foreseeable future (although important progress has been made on computing Mellin moments of double parton distributions on the lattice \cite{Bali:2018nde, Bali:2020mij, Bali:2021gel, Reitinger:2024ulw}, and proposals have been made to directly compute their $x$ dependence \cite{Zhang:2023wea, Jaarsma:2023woo}). Thus for predictions we need to construct models. A large number of model predictions exist in the valence sector \cite{Chang:2012nw, Rinaldi:2014ddl,Rinaldi:2016mlk,Traini:2016jru, Ceccopieri:2017oqe,Rinaldi:2018bsf, Rinaldi:2020ybv, Peng:2024qpw, Dumitru:2025bib}, but since MPI typically probes relatively small $x$ values, account of the sea partons is essential. Monte Carlo Event Generators, that simulate full proton-proton collision events, all have (and need to have) models of general MPI and mPDFs. The simplest model of the mPDFs is the Eikonal model \cite{Butterworth:1996zw, Borozan:2002fk} that is used in \herwig, where it is assumed that the different interactions are independent of one another. The current \pythia model has some important improvements on this picture: the interactions are ordered in energy, from high to low, and modifications are made to the PDFs after each interaction to take account of momentum and valence number constraints \cite{Sjostrand:2004ef,Sjostrand:2004pf}\footnote{As an aside, we note that the MPI model in the Angantyr model of  pA and \textit{nucleus-nucleus} (AA) collisions~\cite{Bierlich:2018xfw}, as implemented in \pythia, is based upon a modified version of \pythia's MPI model, and can be used to simulate DPS in pA and AA collision systems~\cite{Fedkevych:2019ofc}.}. The MPI model of  \sherpa  is based upon the old MPI model of \pythia~\cite{Sjostrand:1987su} with some modification as described in~Ref.~\cite{Gleisberg:2008ta}.

Although the equal-scale mPDFs are unknown, we know two important constraints on these objects. First, the mPDFs should remain the same when we swap the flavour and momentum indices of parton $i$ in the mPDF with parton $k$, $\{x_i,j_i\} \leftrightarrow \{x_k,j_k\}$ (since this describes the same collection of partons -- the ordering of the partons in the mPDF has no significance). Second, the mPDFs should obey number and momentum sum rules, which formally express the constraints on valence number and momentum in the proton, and are the mPDF analog of the usual PDF sum rules. The sum rules have been written down and shown to hold for the ($\overline{\text{MS}}$) renormalised mPDFs in the case of $m=2$ \cite{Gaunt:2009re, Diehl:2018kgr} and $m=3$ \cite{Fedkevych:2022myf}, but analogous relations should hold for higher $m$.

Given the \pythia mPDFs represent the most sophisticated model that exists for general $m$, an interesting question to ask is whether the \pythia equal-scale mPDFs satisfy these constraints. We addressed this question in a previous publication \cite{Fedkevych:2022myf}. We found that the \pythia model cannot perfectly satisfy these constraints at the same time, which is in part linked to the intrinsic energy ordering in the model. Either one can define asymmetric mPDFs, which are not symmetric in the exchange of parton indices in the mPDF, but which satisfy the sum rules perfectly only when integrating over the last parton. Or, if one naively symmetrizes the mPDFs (which is somewhat akin to what \pythia actually does when simulating equal-scale MPI), the sum rules are not perfectly satisfied. Studying explicitly the double and triple PDF cases (dPDF and tPDF), we observed very severe $\mathcal{O}(100\%)$ violations when the $x$ fractions of the partons that are not integrated over are large, but in some cases also substantial $> 10\%$ violations where the accompanying partons have small momentum.

Can one build a better model of the equal-scale mPDFs, that is both symmetric in the parton indices and satisfies the sum rules reasonably well (to within $\mathcal{O}(10\%$)) over the full phase space for the non-integrated partons? This is the aim of the present paper. After a brief review of mPDFs, the \pythia model, and our previous work in Section~\ref{sec:pythia_framework}, we introduce a novel model for constructing mPDFs in Section~\ref{sec:Xord_Defn}. The philosophy is to build on the \pythia model, and make the fewest modifications possible to improve the extent to which the sum rules are satisfied, with these modifications being physically motivated as much as possible. We refer to this model as the ``X-ordered'' mPDFs. We focus in particular on the dPDF and tPDF cases, since double and triple scattering (DPS and TPS) are the cases where (semi-)hard MPI sensitive to number/momentum effects is most likely to be measurable/relevant -- there is already an well-established programme of DPS measurements at hadron colliders (see \textit{e.g.}~Refs.~\cite{AxialFieldSpectrometer:1986dfj, UA2:1991apc, CDF:1993sbj,
CDF:1997lmq,  CDF:1997yfa, D0:2009apj, ATLAS:2016ydt, LHCb:2012aiv,
ATLAS:2013aph, CMS:2013huw, D0:2014vql, D0:2014owy, ATLAS:2014ofp, CMS:2013slh, LHCb:2015wvu, D0:2015dyx, D0:2015rpo, ATLAS:2016rnd, CMS:2017han, CMS:2021ijt, LHCb:2020jse, CMS:2021qsn, CMS:2022pio, CMS:2024wgu, ATLAS:2025bcb}), and activity/interest in TPS measurements is growing \cite{dEnterria:2017yhd, Shao:2019qob, CMS:2021qsn, Maneyro:2024twb}. We give a detailed presentation of the X-ordered construction for the dPDF case in Section~\ref{sec:Xord_dPDF_construction}, and then study the extent to which the sum rules are satisfied by these dPDFs in Section~\ref{sec:sum_rule_assesment_dPDFs}. Section~\ref{sec:tpdfs_results} presents the corresponding exercise for tPDFs. We do not present or study the construction for the mPDF case, but the generalization to this case is fairly straightforward, and we would also expect these to satisfy the sum rules rather well over the full phase space (although looking at the trend between the dPDFs and tPDFs, it likely would get gradually worse as $m$ increases). In section \ref{sec:on_shell_production} we make a toy study of how the changes in the X-ordered dPDFs affect DPS cross sections, focussing on the same-sign $W^\pm W^\pm$ and $ZZ$ cross sections, and also comparing with the GS09 set of dPDFs \cite{Gaunt:2009re}. Finally, we summarize and present some possible future directions to the work in Section~\ref{sec:summary}.

\section{Multiple parton distributions (mPDFs) and the \pythia model}
\label{sec:pythia_framework}

\subsection{Multiple Parton Densities and GS sum rules}
\label{sec:dpdfs_and_sum_rules}
\input{Sections/dPDFs_and_Sum_Rules}

\input{Sections/Modified_mPDFs}

\subsection{Sum Rule assessment of the X-ordered dPDFs}
\input{Sections/DPDF_Sum_Rule_Assessment}
\label{sec:sum_rule_assesment_dPDFs}

\input{Sections/TPD_Sum_Rules}

\section{Toy Phenomenology of the X-ordered and PYTHIA dPDFs}
\label{sec:on_shell_production}
\input{Sections/PhenoDraft_new}

\acknowledgments
The work of OF is supported in part by the US Department of Energy
(DOE) Contract No. DE-AC05-06OR23177, under which Jefferson Science Associates, LLC operates Jefferson Lab,
by the Department of Energy Early Career Award grant DE-
SC0023304, by the Research Council of Finland Project No. 361179, and through the Centre of Excellence in Quark Matter.
OF also would like to thank
Yang-Ting Chien for useful and interesting discussions. 
Some of the simulation is conducted with computing facilities of the Galileo cluster at the Department of Physics and Astronomy of Georgia State University. 
The work of JRG has been supported by the Royal Society through Grant URF\textbackslash{}R1\textbackslash{}201500.

\newpage
\bibliography{journal.bib}

\clearpage

\end{document}

%% file: preamble.tex
\let\originalleft\left
\let\originalright\right
\renewcommand{\left}{\mathopen{}\mathclose\bgroup\originalleft}
\renewcommand{\right}{\aftergroup\egroup\originalright}

\newcommand{\sherpa}{S\protect\scalebox{0.8}{HERPA}\xspace}
\newcommand{\pythia}{P\protect\scalebox{0.8}{YTHIA}\xspace}
\newcommand{\herwig}{H\protect\scalebox{0.8}{ERWIG}\xspace}

\def\beq{\begin{equation}}  
\def\eeq{\end{equation}}
\def\({\left(}
\def\){\right)}
\def\[{\left[}
\def\]{\right]}

%% file: Sections/dPDFs_and_Sum_Rules.tex
The objects of our study will be the double parton distribution functions (dPDFs) and their multi-parton analogues, the multi-parton distribution functions (mPDFs). We shall focus only on the case where the two (or more) partons are probed at the same hard scale $Q$. The dPDFs/mPDFs encode the correlations in longitudinal momentum fractions $x_i$ of the probed partons, with the dependence on transverse space integrated over. More precisely, the equal-scale dPDF $D_{j_1,j_2}$ is defined according to
\begin{equation}
    \begin{aligned}
    D_{j_1,j_2}(x_1,x_2,Q)=\int_{|\bm{y}|>1/Q} \mathrm{d}^2y \, F_{j_1,j_2}(x_1,x_2,\bm{y},Q),
    \label{eq:dPDF_def}
    \end{aligned}
\end{equation}
where the objects $F_{j_1,j_2}$ are the so-called double parton densities (DPDs) \cite{Diehl:2011tt,Diehl:2011yj,Diehl:2017kgu}. Roughly speaking, the DPDs represent the number density of parton pairs with flavours $j_1$ and $j_2$, momentum fractions $x_1$ and $x_2$, separated by transverse distance $\bm{y}$, when the two partons are probed at scale $Q$. To obtain the dPDF, the DPD is integrated over $\bm{y}$, with a lower cutoff inserted to regulate a UV divergence in the DPDs as $\bm{y} \to 0$. One may also regulate the $\bm{y}$ integral using dimensional regularisation and $\overline{\text{MS}}$ renormalisation; if one sets the corresponding renormalisation scale to $Q$ then the resulting dPDFs agree with those defined in \eqref{eq:dPDF_def} up to subleading corrections of order $\alpha_s$ and/or $\Lambda_{\text{QCD}}/Q$. The mPDFs are defined in terms of the $\bm{y}_i$-dependent multiple parton distributions (MPDs) in an analogous way. 

The DPDs/MPDs are the quantities that are actually probed in the DPS/MPI cross sections (see below); nonetheless is it valuable to study the dPDFs/mPDFs since appropriate modeling of these is an important stepping stone towards more realistic modeling of the full DPDs/MPDs (an example of this can be seen in Refs.~\cite{Gaunt:2009re} and \cite{Diehl:2020xyg}; the former paper developed a model of dPDFs, which was later extended into a $\bm{y}$-dependent model in~Ref.~\cite{Diehl:2020xyg}).

The dPDFs obey the momentum and number sum rules \cite{Gaunt:2009re, Gaunt:2012tfk, Blok:2013bpa, Diehl:2018kgr}:
\begin{eqnarray}
	&&\sum\limits_{j_2}\int\limits^{1 - x_1}_0 \mathrm{d}x_2 \, x_2 \, D_{j_1 j_2} (x_1, x_2, Q) = (1 - x_1)\,f_{j_1} (x_1, Q),
	\label{eq:gs_momentum_rule_inv}\\
	&&\int\limits^{1 - x_1}_0 \mathrm{d}x_2 \,  D_{j_1 j_{2v}}(x_1, x_2, Q) = \left( N_{j_{2v}} - \delta_{j_1 j_2} + \delta_{j_1 \bar{j}_2} \right) f_{j_1}(x_1, Q),
	\label{eq:gs_number_rule_inv}
\end{eqnarray}
where $D_{j_1 j_{2v}} = D_{j_1 j_2} - D_{j_1 \bar{j}_2} $. These relations hold at all orders for the $\overline{\text{MS}}$ renormalized dPDFs \cite{Diehl:2018kgr} $-$ for the cut-off regularized dPDFs defined in Eq.~\eqref{eq:dPDF_def} there will be subleading corrections to these relations, which however we shall not concern ourselves with here.

Denoting the mPDF as $\mathcal{F}^{m}$, it is straightforward to write down a proposed form for the mPDF momentum and number sum rules:
\begin{eqnarray}
	\sum\limits_{j_2}\int\limits^{1 - x_1}_0 \mathrm{d}x_m \, x_m \, \mathcal{F}^{m}_{j_1...j_m} (x_1,...,x_m, Q) &=& \left( 1 - \sum_{i=1}^{m-1}x_i\right)\,
    \nonumber \\
    && \times \mathcal{F}^{m-1}_{j_1...j_{m-1}} (x_1,...,x_{m-1}, Q),
	\label{eq:momentum_rule_mPDF}\\
	\int\limits^{1 - x_1}_0 \mathrm{d}x_m \,  \mathcal{F}^{m}_{j_1...j_{mv}}(x_1,...,x_m, Q) &=& \left( N_{j_{mv}} - \sum_{i=1}^{m-1}\left(\delta_{j_i j_m} - \delta_{j_i \bar{j}_m}\right) \right) 
    \nonumber \\
    && \times \mathcal{F}^{m-1}_{j_1...j_{m-1}}(x_1,...,x_{m-1}, Q).
	\label{eq:number_rule_mPDF}
\end{eqnarray}
If one defines $\mathcal{F}^0 = 1$, this general expression also covers the well-known single PDF sum rules (for discussion of these see \textit{e.g.} Ref.~\cite{Collins:2011zzd})). Aside from the single and double PDF cases, the sum rules have also been proven for the case of the triple parton distribution functions (tPDFs, which we shall denote in the below using the letter $T$) \cite{Fedkevych:2022myf}. This was shown at all orders for the $\overline{\text{MS}}$ renormalised tPDFs, similar to what was done for the dPDFs. We shall refer to the sum rules \eqref{eq:momentum_rule_mPDF} and \eqref{eq:number_rule_mPDF} as the \textit{GS sum rules} in this paper.

The scale dependence of the dPDFs is given by the following ``double DGLAP'' evolution equation at leading order
\begin{align}
	\frac{\mathrm{d} D_{j_1 j_2} (x_1, x_2, t)}{\mathrm{d}t} &= 
	\frac{\alpha_s(t)}{2\pi} \sum\limits_{j^\prime_1} \, \int\limits^{1 - x_2}_{x_1} \, \frac{dx^\prime_1}{x^\prime_1} \, 
	P_{j^\prime_1 \rightarrow j_1}\left(\frac{x_1}{x^\prime_1}\right)  \, D_{j^\prime_1 j_2}(x^\prime_1, x_2, t) \nonumber\\
	&\phantomrel{=}+\,\frac{\alpha_s(t)}{2\pi} \sum\limits_{j^\prime_2} \, \int\limits^{1 - x_1}_{x_2} \, \frac{dx^\prime_2}{x^\prime_2} \, 
	P_{j^\prime_2 \rightarrow j_2}\left(\frac{x_2}{x^\prime_2}\right) \, D_{j_2 j^\prime_2}(x_1, x^\prime_2, t) \nonumber\\
	&\phantomrel{=}+\,\frac{\alpha_s(t)}{2\pi} \,\sum\limits_{j^\prime} \, 
	P_{j^\prime \rightarrow j_1 j_2}\left(\frac{x_1}{x_1 + x_2}\right) \,f_{j^\prime}(x_1 + x_2, t) \, \frac{1}{x_1 + x_2},
	\label{eq:double_dglap}
\end{align}
with $t = \log(Q^2 / Q^2_0)$, $P_{i\to j}$ the usual DGLAP splitting functions, and $P_{i\to jk}$ the ``$1 \to 2$'' splitting functions (at leading order these are related to the usual DGLAP splitting functions in a straightforward way -- see \textit{e.g}.~Ref.~\cite{Gaunt:2009re}). We refer to the final inhomogeneous term as the ``$1 \to 2$'', ``splitting'' or ``single PDF feed'' term. If the dPDFs obey the sum rules \eqref{eq:gs_momentum_rule_inv} and \eqref{eq:gs_number_rule_inv} at some scale $Q_0$, then evolution according to the double DGLAP equation preserves the sum rules at higher scales \cite{Gaunt:2009re} -- this is shown explicitly for the case of the momentum sum rule in Appendix A of Ref.~\cite{Blok:2013bpa}, where the role of the splitting term of the DGLAP equation in maintaining the sum rule is emphasised. For the case of the tPDFs, the form of the leading order evolution equations is given in Ref.~\cite{Snigirev:2016uaq}.

As mentioned above, the cross section for DPS is expressed in terms of the DPDs
\begin{equation}
    \begin{aligned}
    \sigma^{DPS}_{A,B}=&\frac{1}{1+\delta_{AB}} \sum_{j_1,j_2,j_3,j_4} \int_{y>1/Q}\prod_{i=1}^2 \mathrm{d}x_{i}\mathrm{d}\bar{x}_{i}\mathrm{d}^2\bm{y}\,F_{j_1,j_2}(x_1,x_2,\bm{y},Q) \\
    &\times F_{j_3,j_4}(\bar{x}_1,\bar{x}_2,\mathbf{y},Q)\,
    \hat{\sigma}_{j_1,j_3\rightarrow A}\hat{\sigma}_{j_2,j_4\rightarrow B},
    \label{eq:dPDF_Scatter}
    \end{aligned}
\end{equation}
where this expression is written for the case where $A$ and $B$ have the same energy scale, $Q_A = Q_B = Q$, and $\delta_{AB}$ is a symmetry factor that is $1$ if $A=B$ and $0$ otherwise.

If one makes the assumption that DPDs can be factorised into a dPDF and a smooth flavour- and scale-independent transverse profile with width of approximately the proton radius:
\begin{equation}
    \begin{aligned}
    F_{j_1,j_2/h}(x_1,x_2,\bm{y},Q)\approx D_{j_1,j_2/h}(x_1,x_2,Q)
    \,G(\bm{y}),
    \end{aligned}
    \label{eq:DPDfactorised}
\end{equation}
then the DPS cross section can be rewritten in terms of the  dPDFs 
\begin{equation}
    \begin{aligned}
    \sigma^{DPS}_{A,B}=&\frac{1}{1+\delta_{AB}} \frac{1}{\sigma_{\rm eff}}\sum_{j_1,j_2,j_3,j_4} \int\prod_{i=1}^2 \mathrm{d}x_{i}\mathrm{d}\bar{x}_{i}\,D_{j_1,j_2}(x_1,x_2,Q) \\
    &\times D_{j_3,j_4}(\bar{x}_1,\bar{x}_2,Q)\,
    \hat{\sigma}_{j_1,j_3\rightarrow A}\hat{\sigma}_{j_2,j_4\rightarrow B},
    \label{eq:dPDF_Scatter_noy}
    \end{aligned}
\end{equation}
where $1/\sigma_{\rm eff}$ is simply some overall geometrical prefactor equal to $\int d^2y \, G(\bm{y})^2$. Now, it is known that \eqref{eq:DPDfactorised} cannot hold $-$ in particular, for a ``$1 \to 2$'' splitting contribution to the DPD occurring at the scale $k$, the transverse area associated with such partons should be of order $1/k^2$ rather than the transverse area of the proton. Thus Eq.~\eqref{eq:dPDF_Scatter_noy} does not correctly account for the transverse dependence of the ``$1 \to 2$'' splitting effects and underestimates the impact of these (as well as not accounting for overlap between single and double scattering). Even though this expression for the DPS cross section is an oversimplification, we will use it in the toy phenomenological studies later in this paper to give an indication of the impact of the longitudinal correlations in different dPDF models in the context of cross sections. It is worth also noting that this simplified approach to computing DPS cross sections is effectively what is used by \pythia MC.

\subsection{The \pythia model of multiple parton distributions}
\label{sec:pythia_model}
Here we very briefly summarize the \pythia approach to modeling dPDFs and mPDFs \cite{Sjostrand:2004pf,Sjostrand:2004ef}, and the extent to which these dPDFs and mPDFs satisfy the GS sum rules, as was studied in detail in Ref.~\cite{Fedkevych:2022myf}. 
A key aspect of the model is that multiple scatterings are ``ordered'' in terms of their hardness, from high to low energy (as is typical in a parton shower formulation of an event). The highest-energy ``first'' scattering is simulated according to the unmodified ``raw'' PDFs, and then for the ``subsequent'' lower energy scatterings, the PDFs are modified to take account of the momentum and quark flavour that has been removed out of the proton by previous scatterings. There are four key modifications that are made to achieve this. We briefly outline these below, where we will omit the dependence of the functions on the scale $Q$ for the sake of notational simplicity.
\begin{enumerate}
    \item \textit{``Momentum squeezing''}: At the $n^{\rm th}$ interaction, the momentum available for further interactions is given by $X_n \equiv 1 - \sum_{i=1}^{n-1} x_i$, where $x_i$ is the momentum of the parton involved in the $i^{\rm th}$ interaction. All PDFs are squeezed by $X_n$ as follows
    \begin{equation}
    f_{j_n}(x_n) \rightarrow \frac{1}{X_n}f_{j_n}(x_n/X_n).
    \end{equation}
    This transformation ensures that the PDFs for the 
    $n^{\rm th}$ interaction lie within $0 \le x \le X_n$, and the momentum sum integral applied for to these PDFs yields $X_n$.
    \item \textit{``Valence number subtraction''}: The valence quark distributions for the $n^{\rm th}$ interaction are rescaled to take account of valence quarks removed in previous interactions
\begin{equation}
q_{fvn}(x)=\frac{N_{fvn}}{N_{fvo}}\frac{1}{X_n}q_{fv0}(x/X_n),
\end{equation}
    where $N_{fvn}$ is the number of valence quarks remaining at the $n^{\rm th}$ interaction, $N_{fv0}$ the initial number, and $q_{fv0}$ the unmodified valence quark distribution. 
    \item \textit{``Companion quark addition''}: If a sea quark participates in an interaction, then a ``companion'' contribution is added to the corresponding antiquark distribution, which accounts for the fact that the ``removal'' of the sea quark leads to a flavour imbalance in the remaining sea. If a sea quark with momentum fraction $x_s$ is removed in the first interaction, then the form of the companion distribution added at the second interaction is
\begin{equation}
q_{c0}(x_c,x_s)=\mathcal{C}(x_s)\mathcal{P}_{g\rightarrow q\bar{q}}\left(\frac{x_s}{x_c+x_s}\right)\frac{g(x_s+x_c)}{x_s+x_c}.
\label{Pythia_companion}
\end{equation}
    The quantity $\mathcal{C}(x_s)$ is chosen such that the number integral of $q_c$ is equal to $1$. The momentum squeezing applied to the companion quark distribution at interaction $n$ is given by
\begin{align}
    q_{c}(x_c) = \dfrac{1}{X_n+x_s}q_{c0}\left(\dfrac{x_c}{X_n+x_s},\dfrac{x_s}{X_n+x_s}\right), 
\end{align}
    which ensures that $x_c < X_n$.
    \item \textit{``Sea quark and gluon rescaling''}: The second and third modifications, which are introduced according to quark number considerations, alter the momentum of the proton such that the momentum integral is no longer $X_n$. To repair this, the (non-companion) sea quark and gluon distributions are adjusted by a factor $a$, defined according to
    \begin{equation}
    a=\dfrac{1 - \sum_f\left(N_{fvn}\langle x_{fv0}\rangle - \sum_j\langle x_{fc_{j}0}\rangle\right) }{1-\sum_f N_{fv0}\langle x_{fv0}\rangle}
    \label{eq:adefPythia}
   \end{equation}
with\footnote{Note that our expression for $\langle x_{fc_{j}0}\rangle$ differs from that presented in Ref.~\cite{Sjostrand:2004pf}; the expression here is the one that is actually used in the \pythia code, where it is also noted that there is a typo in the equation given in the original publication, Ref.~\cite{Sjostrand:2004pf}.}
\begin{align}
\langle x_{fc_{j}0}\rangle &= \left( 1 + \dfrac{x_j}{X_n}\right)\int \mathrm{d}x_c \, x_c \, q_{c0}\left(x_c, \dfrac{x_j}{X_n + x_j}\right), \\
\langle x_{fv0}\rangle &= \int \mathrm{d}x \, x \, q_{v0}(x).
\end{align}
    In the sum over $j$ in \eqref{eq:adefPythia}, one only picks up contributions from interactions where a sea quark is selected.
\end{enumerate}

If one constructs a dPDF according to this model, then the GS sum rules are exactly satisfied when integrating over the second parton, and the same statement holds true if one constructs a tPDF and integrates over the third parton \cite{Fedkevych:2022myf}. However, in the equal-scale case, the concept of ``first'', ``second'' or ``third'' interactions is not meaningful, and the mPDF should be symmetric when swapping over the flavours and momentum fractions of any of the partons (and consequently should satisfy the GS sum rules when integrating over any of the partons). One can make a naive symmetrization of the \pythia mPDFs $-$ \textit{e.g.} for the dPDF
\begin{eqnarray}
    D^{\rm sym}_{j_1,j_2}(x_1,x_2) &=& 
    \frac{D^{\rm pth}_{j_1,j_2}(x_1,x_2) + 
    D^{\rm pth}_{j_2,j_1}(x_2, x_1)}{2}.
\end{eqnarray}
This kind of symmetrization is close to what \pythia actually does when simulating equal-scale DPS events, although the symmetrization is done at the level of the luminosity rather than at the level of the dPDFs. In Ref.~\cite{Fedkevych:2022myf} it was found that the dPDFs and tPDFs symmetrized in this simple fashion satsify the sum rules reasonably well, although there are violations of the sum rule beyond the 10-20\% level in certain regions of phase space (and in places violation of the order of 100\% or larger). In Table~\ref{UUSymmtable} we present some sum rule values obtained using the symmetrised \pythia dPDFs. The numbers are obtained by integrating the relevant number and momentum response function over $x_2$, after multiplying by $x_2$. The number and momentum response functions were introduced in Ref.~\cite{Fedkevych:2022myf} and are defined respectively by:
\begin{align}
    R_{j_1j_2}(x_1,x_2) &\equiv x_2 \dfrac{D_{j_1j_2}(x_1,x_2)-D_{j_1\bar{j}_2}(x_1,x_2)}{f_{j_1}(x_1)},
    \\
    R_{j_1}(x_1,x_2) &\equiv  \sum_{j_2}\dfrac{D_{j_1j_2}(x_1,x_2)}{(1-x_1)f_{j_1}(x_1)},
\end{align}
and are constructed such that if the sum rules are perfectly satisfied then we have:
\begin{align}
    \int dx_2 x_2 R_{j_1j_2}(x_1,x_2) &= \left( N_{j_{2v}} - \delta_{j_1j_2} + \delta_{\bar{j}_1j_2}\right), \label{eq:responsenumberint}
    \\
    \int dx_2 x_2 R_{j_1}(x_1,x_2) &= 1.\label{eq:responsemomentumint}
\end{align}
In Table~\ref{UUSymmtable}, and in the rest of the paper, we use single PDFs that essentially correspond to the leading order MMHT 2014 PDFs~\cite{Harland-Lang:2014zoa}. The only slight modification we make from this set is that we  enforce that the strange quark distribution is equal to the strange antiquark distribution
\begin{equation}
 f_{s}(x) = f_{\bar{s}}(x) = \dfrac{f_{s,\text{MMHT2014}}(x) + f_{\bar{s},\text{MMHT2014}}(x)}{2}.
\end{equation}
This is to avoid dealing with issues associated with the numerically small ``valence'' strange quark PDF -- a similar approach was taken in Ref.~\cite{Gaunt:2009re}. The algorithm from the MSTW code \cite{MSTWweb} is used to extrapolate these PDFs below $x = 10^{-6}$ where needed. We also compute $\langle x_{fv0}\rangle$ directly from the PDFs, whilst the actual \pythia code uses a hardcoded parameterization computed from the CTEQ5L PDFs \cite{Lai:1999wy}.

In Table~\ref{UUSymmtable}, one observes some substantial violations of the sum rules, particularly towards large $x$. Note that the numbers given in the table differ slightly from those given in Ref.~\cite{Fedkevych:2022myf} due to the differences in setup.

\begin{table}[t]
\centering
\begin{tabu}{ |p{3cm}||p{4cm}|p{4cm}|}
 \hline
 \multicolumn{3}{|c|}{Symmetrised Model GS Rules} \\
 \hline
$x_1$ &$uu_v$ Number Sum Rule & $u$ Momentum Sum Rule\\
 \hline
 $10^{-6}$   & 0.907&0.975\\
 $10^{-3}$ &  0.836& 0.973\\
 $0.1$ & 0.917& 1.013\\
 $0.2$  & 0.921& 1.047\\
 $0.4$&   0.881& 1.133\\
 $0.8$& 0.748& 1.673\\
 \tabucline[2pt]{-}
 Expected & 1 & 1 \\
 \hline
\end{tabu}
\caption{Results of numerical integration of the symmetrized \pythia dPDF distribution, for the GS number sum rule \eqref{eq:gs_number_rule_inv}  and momentum sum rule \eqref{eq:gs_momentum_rule_inv}, evaluated at $Q = M_Z$. }
\label{UUSymmtable}
\end{table}

%% file: Sections/Modified_mPDFs.tex
\section{``$X$-Ordering'' and the modified \pythia mPDF model}
\label{sec:Xord_Defn}

\subsection{Construction of the model for the dPDF case}
\label{sec:Xord_dPDF_construction}

In this section we will make some improvements to the \pythia mPDF model, yielding a set of symmetric mPDFs that satisfy the GS sum rules to a better extent than the naively symmetrised \pythia mPDFs. We first discuss the simplest case of the dPDFs, postponing the generalization of the new model to the case of tPDFs and more general mPDFs to the end of this section. Our philosophy is to make modifications that are physically motivated if possible, and that are as simple as possible, such that they can be straightforwardly generalized to the mPDF case. 

We also set a (somewhat arbitrary) target of satisfying the sum rules over the full range of $x_1$ and all flavour indices at better than $10\%$ precision. We define the percentage deviation from the sum rule expectation by taking the appropriate response function integral on the left hand sides of Eqs.~\eqref{eq:responsenumberint} or \eqref{eq:responsemomentumint}, dividing by the expected right hand side, and then subtracting $1$. This is unless the expected value is 0, in which case we simply use the integral of the response function to define the percentage deviation.

Let us consider first the modeling of the dPDF for a sea quark and its corresponding companion quark (in the following we  call this the ``sea pair'' dPDF). In \pythia the dPDF for this pair is given according to
\begin{eqnarray}
D_{j_1\bar{j}_1}^{\rm com,pth}(x_1,x_2)&=&
f^s_{j_1}(x_1)q_c(x_2,x_1)\nonumber\\
&=&f^s_{j_1}(x_1)\mathcal{C}(x_1)\mathcal{P}_{g\rightarrow q\bar{q}}\left(\frac{x_1}{x_2+x_1}\right)\frac{g(x_1+x_2)}{x_1+x_2}.
\label{PYTH_comp_dpdf}
\end{eqnarray}
If one wants to imagine the sea quark and its companion as essentially arising from the splitting of a gluon into a quark-antiquark pair (similarly to the way it is done in the \pythia code), one notices some deficiencies in the model when the dPDF is written out explicitly like this. First, in the gluon splitting picture one expects a complete symmetry in the distribution of the momentum fractions for the quark and its companion, whereas this is clearly not the case for \eqref{PYTH_comp_dpdf}. Second, there is a kind of ``double counting'' of the $g \to q\bar{q}$ splitting process in \eqref{PYTH_comp_dpdf}. It appears in the generation of the sea quark distribution $f^s_{j_1}(x_1)$ (recall that at low $x$, the sea quark distribution is mainly sourced from the gluon distribution via a single $g \to q\bar{q}$ splitting process, see e.g. Ref.~\cite{Ralston:1986hr}), and explicitly in the companion quark distribution.

In order to resolve these issues we can remove $f^s_{j_1}(x_1)$ from the definition of the sea pair dPDF, and then take $\mathcal{C}$ to be a function of $x_1+x_2$ rather than $x_1$, such that the dPDF is explicitly symmetric in $x_1$ and $x_2$. Since the $g \to q\bar{q}$ splitting function is not a strong function of the splitting variable (it varies only between $1/2$ and $1$), let us also approximate this as a constant, absorbing this constant into the definition of $\mathcal{C}$. Then we get our new model for the ``sea pair'' dPDF
\begin{eqnarray}
D_{j_1\bar{j}_1}^{{\rm com},m}(x_1,x_2)=\mathcal{C}(x_1+x_2)\frac{g(x_1+x_2)}{x_1+x_2}.
\label{comp_dpdf_1}
\end{eqnarray}
From this definition one can define a \pythia-style companion distribution which is simply this dPDF divided by $f^s_{j_1}(x_1)$. Imposing that the number integral over this companion distribution should be equal to $1$ and solving for $\mathcal{C}$, one obtains the following form for the new ``sea pair'' dPDF and corresponding companion distribution:
\begin{eqnarray}
    D_{j_1j_2}^{{\rm com}, m}(x_1,x_2)&=&-\left[\partial_{u}  \left. f^s_{j_1}(u)\right]\right|_{u=x_1+x_2}, \\
    q_c(x_2,x_1)&=& 
    -\frac{\left[\partial_{u} \left. f^s_{j_1}(u)\right]\right|_{u=x_1+x_2}}{{f^s_{j_1}(x_1)}}. \label{eq:newcompanion}
\label{Xord_comp}
\end{eqnarray}
Note that this is essentially the same sequence of steps as is done in the modeling of the ``$j\bar{j}$ correlation term'' in Ref.~\cite{Gaunt:2009re}, where that term plays a similar role in the modeling of the GS09 input distributions in that paper as the ``sea pair'' dPDF does here. 

We now have a companion quark mechanism that functions to account for the imbalance in the sea when a quark is removed, as before, but now is explicitly symmetric in the momentum fractions of the sea quark and its companion.

Let us recall from the end of Section~\ref{sec:pythia_model} that the region where the sum rules were generally most poorly satisfied by the symmetrized \pythia dPDFs was the region of large $x_1 \gtrsim 0.4$. The sum rules in this region could be significantly improved if, rather than simply symmetrizing the \pythia dPDFs in a naive way, we impose an $x$-ordering procedure, enforcing that the parton with larger $x$ be picked ``first''
\begin{equation}
D_{j_1j_2}^{\{1,2\}}(x_1,x_2) \equiv \theta(x_1 - x_2) D^{\rm{pth}}_{j_1j_2}(x_1,x_2) + \theta(x_2 - x_1) D^{\rm{pth}}_{j_2j_1}(x_2,x_1).
\label{eq:Xordsimple}
\end{equation}
Note that this dPDF is symmetric under the interchange $\{x_1,j_1\} \leftrightarrow \{x_2,j_2\}$ as required. It also perfectly satisfies the sum rules by construction for $x_1 > 0.5$, since in this region $x_1$ must be greater than $x_2$, parton $1$ is then always forced to be first, and the \pythia algorithm always satisfies the sum rule perfectly in the second parton. For values of $x_1$ that are large but $< 0.5$, the sum rules should still be satisfied well, since over most of the $x_2$ integration range we will have $x_2 < x_1$. 

From an intuitive point of view, this ``X-ordering'' procedure seems at least somewhat reasonable. Let us say that we have the situation where a large-$x$ parton and a small-$x$ parton ($x_1 \gg x_2$) are involved. It seems plausible to expect that the small-$x$ parton will be affected more by the presence of the large-$x$ parton than vice versa, so we choose to apply the \pythia modifications to the small $x$ parton.

A drawback of the form~\eqref{eq:Xordsimple} is that it has a discontinuity at $x_1=x_2$, since we abruptly switch from parton $1$ being the first, to parton $2$. Rather than having such a discontinuity, we instead choose to implement a smooth transition between using $D_{j_1j_2}^{\{1,2\}}$ in the strongly ordered regime ($x_1 \gg x_2$ or vice versa), and the naive symmetrized dPDF when $x_1 \simeq x_2$. This can be done by introducing a smoothing function $F$ as follows
\begin{equation}
    D_{j_1j_2}(x_1,x_2) = D_{j_1j_2}^{\{1,2\}}(x_1,x_2) \left[ 1 - \dfrac{1}{2}F\left(\dfrac{x_{\rm max}}{x_{\rm min}}\right) \right]+D_{j_1j_2}^{\{2,1\}}(x_1,x_2) \dfrac{1}{2}F\left(\dfrac{x_{\rm max}}{x_{\rm min}}\right),
    \label{eq:smoothed_xord}
\end{equation}
where $x_{\rm max} = \max(x_1,x_2)$, $x_{\rm min} = \min(x_1,x_2)$, and $D_{j_1j_2}^{\{2,1\}}$ is the dPDF computed according to the \pythia procedure, but with the lower $x$ parton always picked ``first''.

The function $F(x)$ is only needed for $x\ge1$. To have the required behaviour, it should take the value of $1$ at $x=1$, and then monotonically decrease to zero as $x \to \infty$. We chose to use the simple form
\begin{equation}
    F(x) = e^{-(x-1)^2}.
    \label{eq:smoothingfunc}
\end{equation}
This procedure improves the extent to which the sum rules at satisfied at very large $x_1$, but may degrade the extent to which they are satisfied at low $x_1$ (at low $x_1$ values, we are probing the ``wrong'' ordering of the dPDFs over most of the $x_2$ integral). We investigated this by computing the extent to which our modified dPDFs satisfy the sum rules, for $x_1 = \{10^{-6}, 10^{-3}$, 0.1, 0.2, 0.4, 0.8\} and \mbox{$Q = \{ 2, 5, 10, 20, 40, M_W, M_Z$ and $100\}$ GeV} (where we take $M_W = 80.385$ GeV, $M_Z = 91.118$ GeV here). At large $Q \sim M_Z$ we found that it was the quark number sum rules in particular that are degraded at small $x_1$ -- see Table~\ref{XordNStable} which contain results for some of the number sum rules at $Q = M_Z$. Even though the sum rules are not violated particularly drastically at small $x_1$, the violation does exceed the 10\% level for some cases. At the low $Q$ values, both gluon and quark number sum rules are violated more severely at small $x_1$ values -- see Table~\ref{XordNS2GeVgluontable}. On the other hand, the momentum sum rules were found to be obeyed within the 10\% tolerance at all $Q$ values.

\begin{table}
\centering
\begin{tabu}{|p{2cm}|p{3cm}|p{3cm}|p{3cm}|}
 \hline
$x_1$ &$uu_v$ Sum Rule& $gu_v$ Sum Rule& $\bar{s}u_v$ Sum Rule\\
 \hline
 $10^{-6}$    & 1.137&2.128&2.134\\ \hline 
 $10^{-3}$    &  1.085&2.065&2.087\\ \hline 
 $0.1$      & 1.003&1.913&1.910\\ \hline 
 $0.2$      & 0.996&1.930&1.925\\ \hline 
 $0.4$      & 0.994&1.969&1.967\\ \hline 
 $0.8$      & 0.997&1.994&1.994\\
 \tabucline[2pt]{-}
 Expected & 1 & 2 & 2 \\
 \hline
\end{tabu}
\caption{Results for selected number sum rules for dPDFs constructed at $Q=M_Z$, with only the ``$X$-Ordering'' mechanism and the symmetric sea pair dPDF (no damping applied). The numbers are the integrals of the response functions, and the expected values if the sum rules are perfectly satisfied are given at the bottom of the table.}
\label{XordNStable}
\end{table}
\begin{table}
\centering
\begin{tabu}{|p{2cm}|p{3cm}|p{3cm}|p{3cm}|}
 \hline
$x_1$ &$uu_v$ Sum Rule& $gu_v$ Sum Rule& $\bar{s}u_v$ Sum Rule\\
 \hline
 $10^{-6}$    & 1.337& 2.302& 2.326\\ \hline 
 $10^{-3}$    & 1.245& 2.257& 2.293\\ \hline 
 $0.1$      & 1.049& 1.971& 1.890\\ \hline 
 $0.2$      & 1.014& 1.982& 1.879\\ \hline 
 $0.4$      & 0.988& 1.949& 1.930\\ \hline 
 $0.8$      & 0.998& 1.996& 1.996\\
 \tabucline[2pt]{-}
 Expected & 1 & 2 & 2 \\
 \hline
\end{tabu}
\caption{The same quantities as in Table \ref{XordNStable}, but now computed for dPDFs constructed at $Q=2$ GeV.}
\label{XordNS2GeVgluontable}
\end{table}

The violation of these number sum rules can be associated with an overcontribution from pairs of sea quarks (plus overcontributions from sea quark-gluon pairs at small $Q$) when both $x_1$ and $x_2$ are small. This suggests that a way to fix these issues is to add a mild damping of these dPDFs when both $x_1$ and $x_2$ are small, where a simple way to do this is to multiply by $(x_2+x_2)^k$. 
We want to apply the damping only where it is strictly needed, and keep the damping as mild as possible -- one reason for this is that whilst increasing amount of damping improves the number sum rules (up to a point), it takes momentum out of the dPDFs and degrades the momentum sum rules. 
The $k$ values for sea quark pair gluon-quark pairs should be different, and both should be dependent on $Q$, with the latter becoming much smaller than the former at large $Q$.
Finally, we do not want to apply this damping factor to our ``sea quark pair'' dPDFs $D^{\text{com},m}$, since these have been constructed to exactly satisfy their relevant ``number sum rule'' (\textit{i.e.} that there is exactly one companion accompanying the sea quark). 
We only apply the damping to the rest of the dPDF, which we shall denote by $D_{j_1j_2}^{v/s}$.
Thus, the form of the dPDFs with the damping factor is
\begin{eqnarray}
    D_{j_1j_2}^{m}(x_1,x_2,Q) = 
    (x_1 + x_2)^{k_{j_1j_2}(Q)} D_{j_1j_2}^{v/s}(x_1,x_2) + 
    \delta_{j_1,\bar{j_1}}D_{j_1j_2}^{{\rm com}, m}(x_1,x_2,Q),
    \label{lowxgdamp}
\end{eqnarray}
where there are two cases in which the damping factor $k_{j_1j_2}$ is nonzero: if $j_1j_2$ is a quark-gluon pair or a gluon-gluon pair, we set $k_{j_1j_2}(Q)=k_g(Q)$, and if $j_1j_2$ is a quark-quark pair that is not a pair of strange quarks, we set $k_{j_1j_2}(Q)=k_q(Q)$. Appropriate forms for $k_{q,g}(Q)$ were established by manually tuning the values of these over the set of $Q$ values discussed above, and then fitting simple functions to the result. We obtain:
\begin{eqnarray}
    k_q(Q)&=&k_{q_sq'_s}(Q)=\frac{0.053}{Q/\text{GeV}}+0.007,\\
    k_g(Q)&=&k_{gj_2}(Q) = k_{j_1g}(Q)=\frac{0.02}{(Q/\text{GeV})^{0.75}-0.5}.
    \label{alphaforms}
\end{eqnarray}

Since these functions have been obtained from data only in the range $2\text{ GeV}<Q<100\text{ GeV}$, care should be taken if one uses them outside this range. In particular, these functions will not give dPDFs satisfying the sum rules to within $10\%$ below $2$ GeV. In fact even at $2$ GeV we were not quite able to satisfy all sum rules to within 10\% everywhere, with some small localized deviations above 10\%, see Section~\ref{sec:sum_rule_assesment_dPDFs}  (whilst at the higher $Q$ values we were able to tune $k_{q,g}$ to achieve better than $10\%$ agreement). For $Q$ values below 2 GeV, it does not appear to be possible to get dPDFs satisfying the sum rules to a satisfactory precision using the simple prescription described above (even when tuning $k_{q,g})$, and further modifications would presumably be needed.
This final modification concludes the definition of our X-ordered dPDFs.

%% file: Sections/DPDF_Sum_Rule_Assessment.tex
Now let us  study how well our X-ordered dPDFs satisfy the sum rules when constructed at a high scale, $Q = M_Z$, and a low scale, $Q=2$ GeV (the latter is the lowest scale at which our construction gives dPDFs satisfying the sum rules reasonably well). 
The $Q=2$ GeV dPDFs could be used as a low-scale model input in pQCD evolution to higher scales, where the appropriate evolution equation for the dPDFs is the inhomogeneous double DGLAP equation, see
Eq.~\eqref{eq:double_dglap}. 

\begin{figure}
\centering
\includegraphics[width=0.8\textwidth]{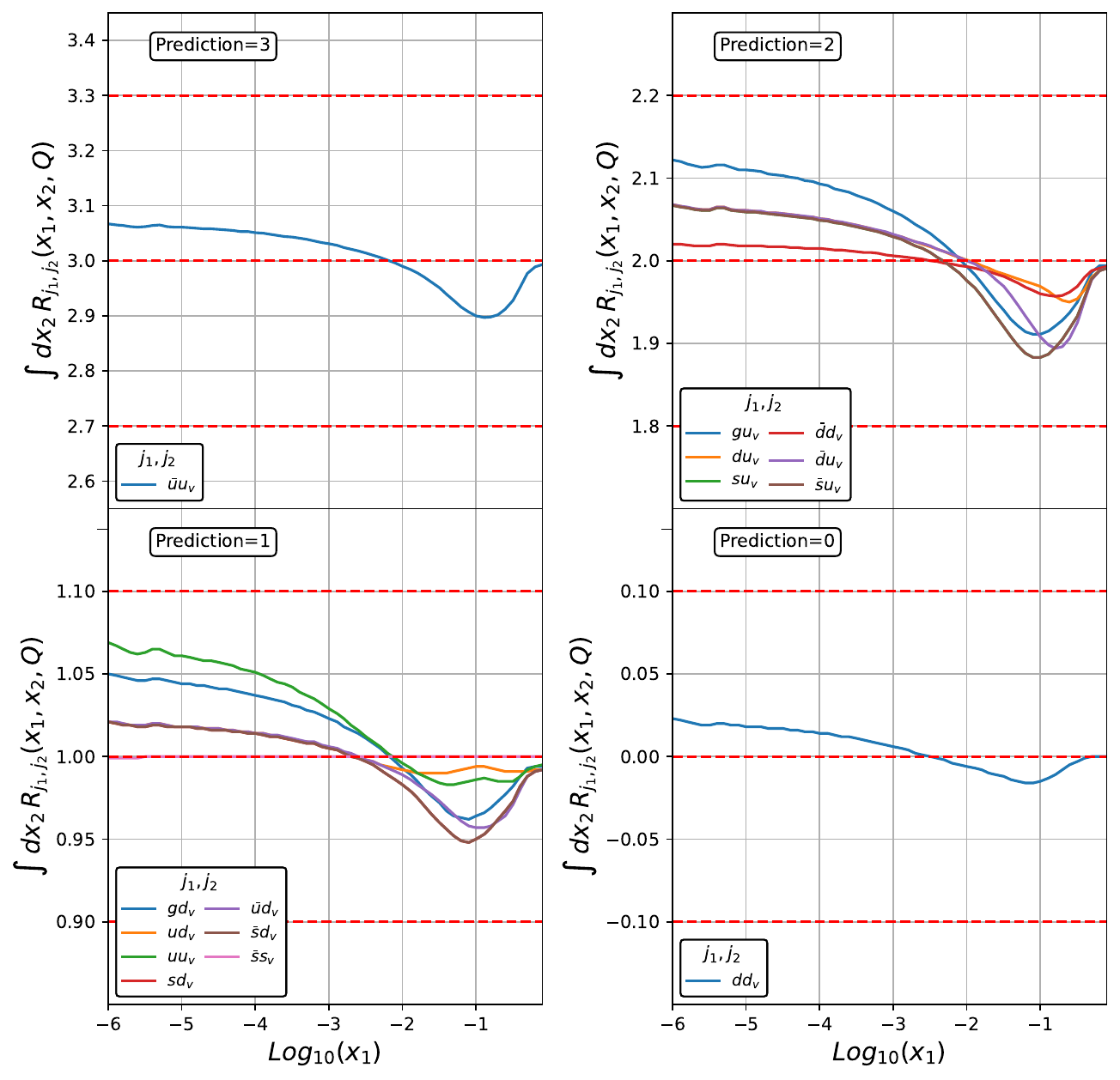}
\caption{Behavior of the Number Sum Rule integrals for all parton flavors in the X-ordered dPDF model at $Q=M_Z$, as defined in Section \ref{sec:Xord_dPDF_construction}.}
\label{XordNSREvals}
\end{figure}

\begin{figure}
\centering
\includegraphics[width=0.6\textwidth]{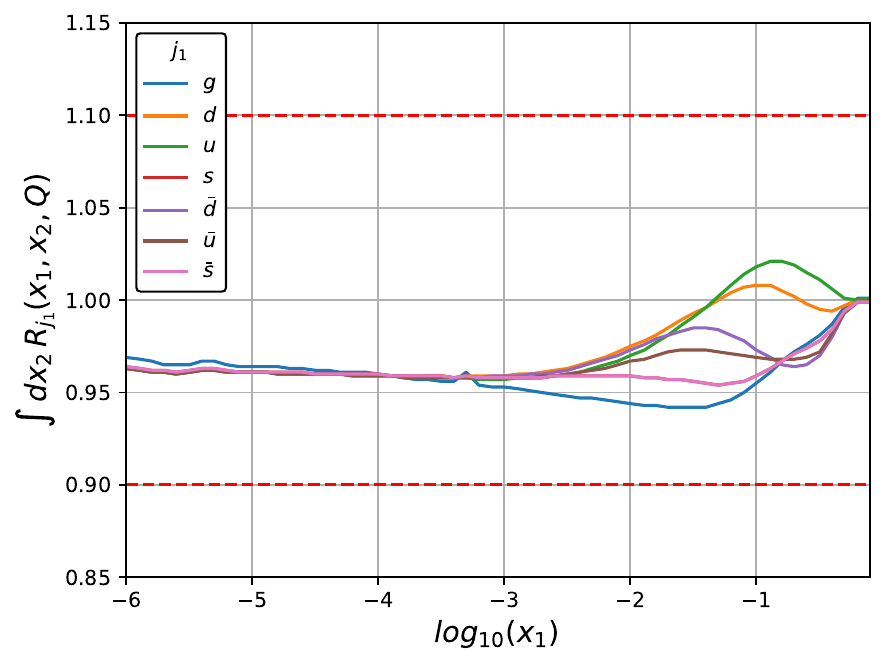}
\caption{Behavior of the Momentum Sum Rule integrals for all parton flavors in the X-ordered dPDF model  at $Q=M_Z$, as defined in Section \ref{sec:Xord_dPDF_construction}.}
\label{XordMSREvals}
\end{figure}

Let us start the discussion by showing the results for the number and momentum sum rules respectively, for X-ordered dPDFs constructed at $Q=M_Z$, see Fig.~\ref{XordNSREvals} and Fig.~\ref{XordMSREvals}. 
The quantities plotted are the integrals of the response functions as defined in \eqref{eq:responsenumberint} and \eqref{eq:responsemomentumint}; the expected value if the sum rule is perfectly satisfied is given by the central red dashed line, with the upper and lower red dashed lines representing a 10\% deviation from expectation. In Fig.~\ref{XordNSREvals} we do not plot the curves for sum rules that are trivially satisfied -- for example any strange number sum rule where the second parton is not a strange or anti-strange will evaluate to zero due to the symmetry of the strange sea quarks. 

We observe that all sum rules are obeyed to well within 10\%, as desired. We also see that the number sum rule curves all have the same qualitative shape, being larger than expectation at small $x_1 \ll 0.1$, dipping below expectation at $x_1 \sim 0.1$, before finally rising to a value just below the expected value as $x_1 \to 1$. The momentum sum rules also have some shared behavior, all being smaller than $1$ for $x_1 \ll 0.1$, and approaching a value just below $1$ for $x_1 \to 1$.

\begin{figure}
\centering
    \includegraphics[width=0.32\textwidth]{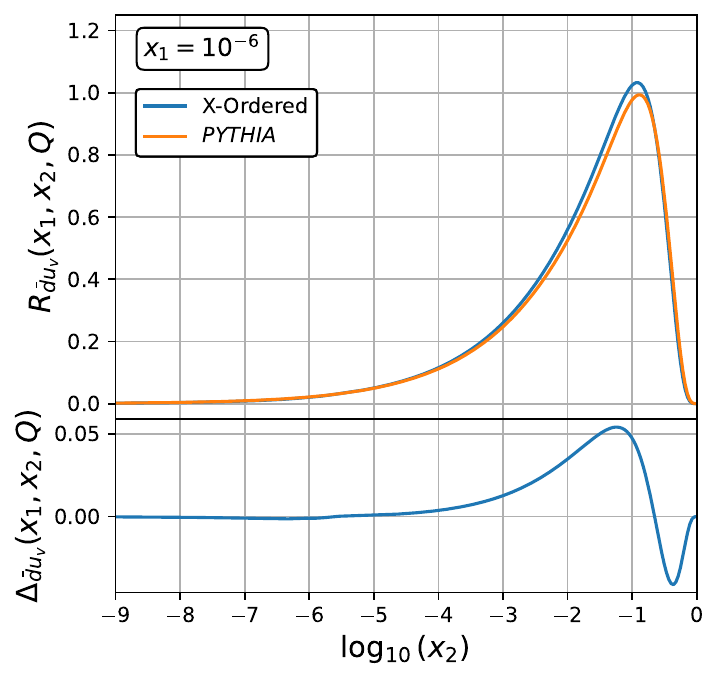}      
    \includegraphics[width=0.32\textwidth]{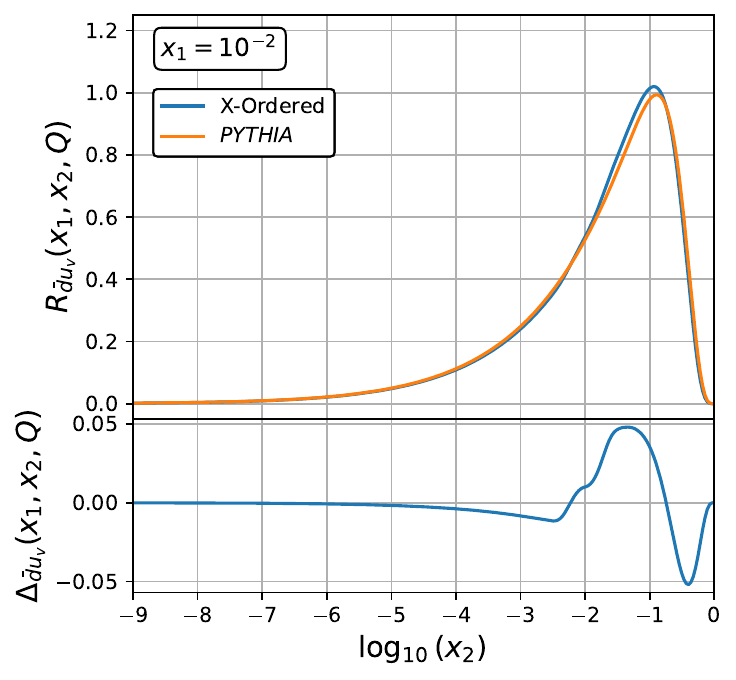}
\includegraphics[width=0.32\textwidth]{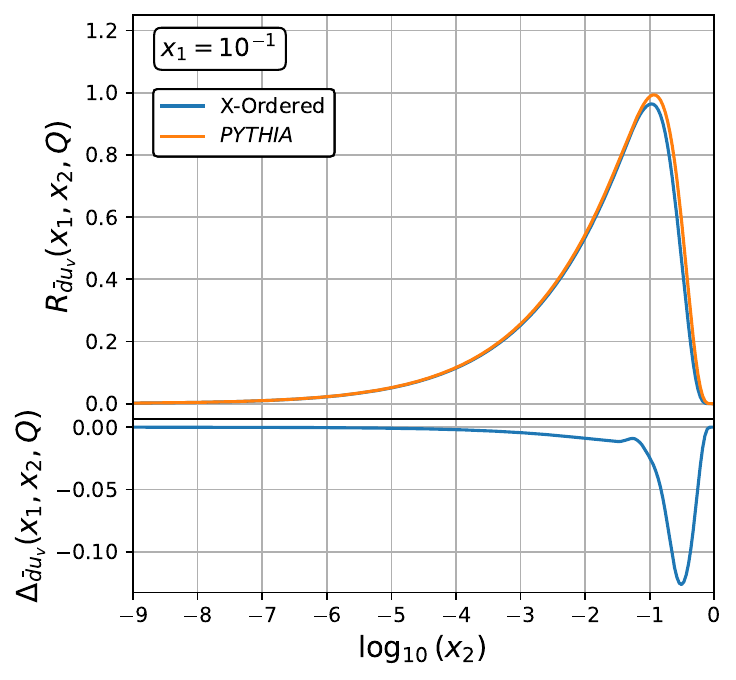}
    \caption{Behavior of the $\bar{d}u_v$ Number Sum Rule response function for our X-ordered dPDFs and for the \pythia (unsymmetrised) dPDFs, at $x_1=\{10^{-6},10^{-2}, 10^{-1}\}$. The bottom panels give the difference between the X-ordered and  \pythia response functions, $\Delta_{\bar{d}u_v}$.}
\label{ubdv_otherx1_plot}
\end{figure}

Let us discuss why the curves have these generic shapes, starting with the number sum rules. We consider the example of the ${\bar{d}u_v}$ case, and to aid the explanation we make plots of the response function curves at several values of $x_1$ in Fig.~\ref{ubdv_otherx1_plot}. In the same figure we draw the response functions for the unsymmetrised \pythia dPDFs, since those response functions integrate to exactly the expected values, and so can be used as reference of what the response function ``should'' look like.

For $x_2 < x_1$ our X-ordered prescription reduces to the \pythia one, apart from the additional damping (and the difference in the companion quark mechanism, which we shall not consider for the moment and is not relevant for the $\bar{d}u_v$ case). Thus, the X-ordered curve lies close to but slightly below the \pythia one, as observed in Fig.~\ref{ubdv_otherx1_plot}. As $x_2$ exceeds $x_1$ our ordering is now ``the wrong way round'', such that for our $\bar{d}u_v$ example our prescription yields
\begin{equation}
    D_{\bar{d}u_v}(x_1,x_2)=
    a_vf_{\bar{d}}\left(\frac{x_1}{1-x_2}\right)\frac{1}{1-x_2}f_{u_v}\left(x_2\right)(x_1+x_2)^{k_q(M_Z)},
    \label{eq:ourduv}
\end{equation}
where $a_v$ is the $a$-factor from Eq.~\eqref{eq:adefPythia} after one up valence quark is removed, whilst for \pythia we have
\begin{equation}
    D^{\mathrm{pth}}_{\bar{d}u_v}(x_1,x_2)=
        f_{\bar{d}}(x_1)f_{u_v}\,
    \left(\frac{x_2}{1-x_1}\right)\frac{1}{1-x_1}.
    \label{eq:pythiaduv}
\end{equation}
If $x_2 \ll 0.1$ the effect of the momentum squeezing in \eqref{eq:ourduv} is rather small. The impact of the momentum squeezing in \eqref{eq:pythiaduv} will be to enhance the \pythia dPDF, although the effect is also somewhat mild unless $x_1 \gtrsim 0.1$. Then, the two prescriptions are rather close up to two additional factors that are present in \eqref{eq:ourduv}: the $a_v$-factor and the damping factor. The $a_v$-factor is $>1$ and outweighs the mild damping factor, pushing the X-ordered response curve above the \pythia one for $x_1 < x_2 \lesssim 0.1$ as seen in 
Fig.~\ref{ubdv_otherx1_plot}. 

For large $x_2 \gtrsim 0.1$, the momentum squeezing is the dominant effect controlling the overall suppression of the dPDF and ensuring it decreases to zero as $x_2 \to 1-x_1$. In Eq.~\eqref{eq:pythiaduv} this overall decrease is set by the $u_v$ valence PDF behavior near $x=1$, whilst in Eq.~\eqref{eq:ourduv} it is set by the $\bar{d}$ sea quark; since sea quark distributions decrease to zero faster than valence ones near $x=1$, this results in the  X-ordered response curve decreasing below the \pythia one for $x_2 \gtrsim 0.1$.

This then explains the behavior we see in Fig.~\ref{XordNSREvals}. For small $x_1 \ll 0.1$ the phase space over which the response function enhancement applies, $x_1 < x_2 \lesssim 0.1$, is larger than that over which there is a suppression, and we see an overall enhancement over the expectation. Then as $x_1$ increases the ``enhanced'' portion of the phase space grows smaller and eventually disappears, leaving a suppression. Finally, when $x_1$ gets larger than 0.5 the ``suppressed'' phase space disappears too, and we end up with a sum rule integral that agrees with expectation, up to a small effect from the damping factor. 

For the momentum sum rule, the picture is somewhat similar, although in this case the significant $a_v$ enhancement will be absent, and the importance of the large $x_2$ region will be increased by the extra factor of $x_2$ in the momentum sum rule integral. Thus, it is not surprising that we now observe an overall suppression at small $x_2$ rather than an enhancement. 

Before moving on, let us briefly touch on the impact of the companion quark mechanism. As mentioned previously, our companion quark mechanism is constructed to perfectly satisfy its relevant sum rule, so it is not responsible for any of the fluctuations away from the expected values in Figs.~\ref{XordNSREvals} and \ref{XordMSREvals}. It is nonetheless interesting to study the effect of our companion quark on number sum response functions, and compare with (unsymmetrised) \pythia. In both cases, the companion introduces a ``bump'' (or dip) in the response function with the same area (since in both cases the companion mechanism satisfies the sum rule when integrating over the second parton), but the shape of the bump or dip can be somewhat different.

\begin{figure}
\centering
\begin{subfigure}
    \centering
    \includegraphics[width=0.32\textwidth]{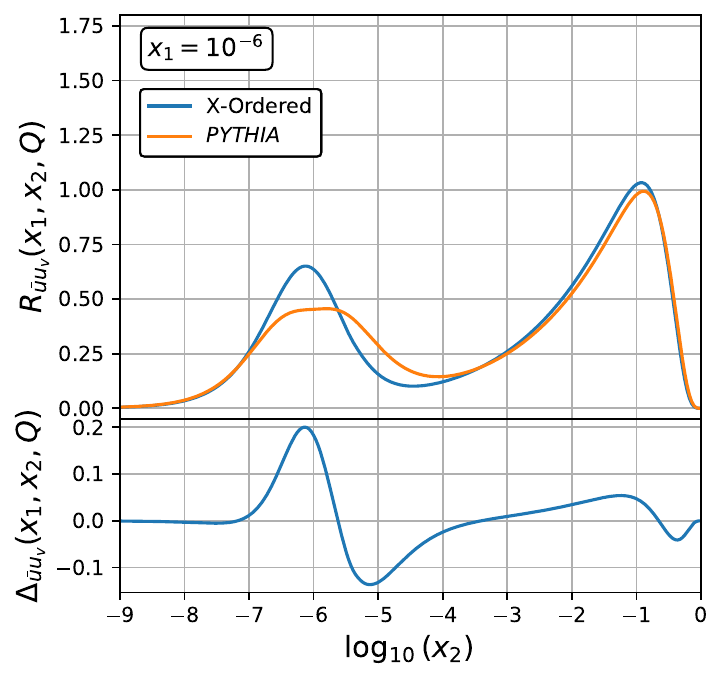}
\end{subfigure}
\begin{subfigure}
    \centering
    \includegraphics[width=0.32\textwidth]{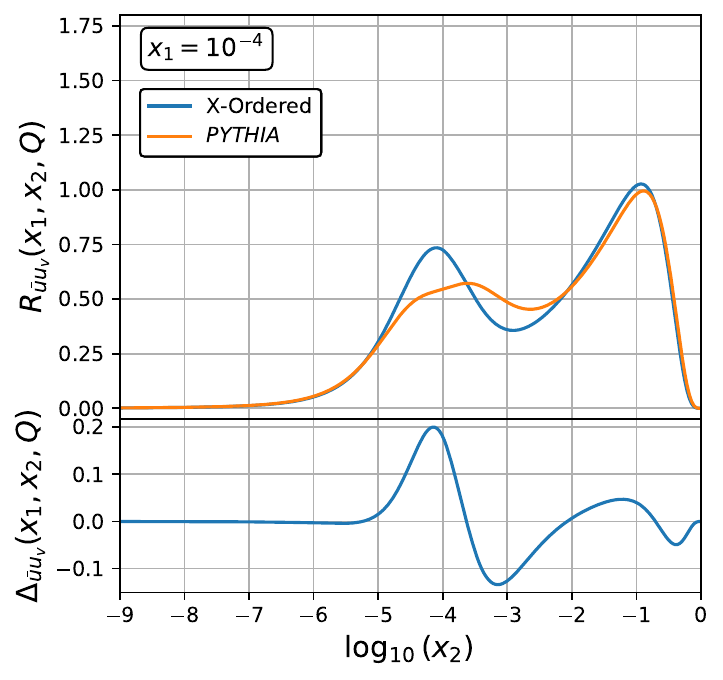}
\end{subfigure}
\begin{subfigure}
    \centering
    \includegraphics[width=0.32\textwidth]{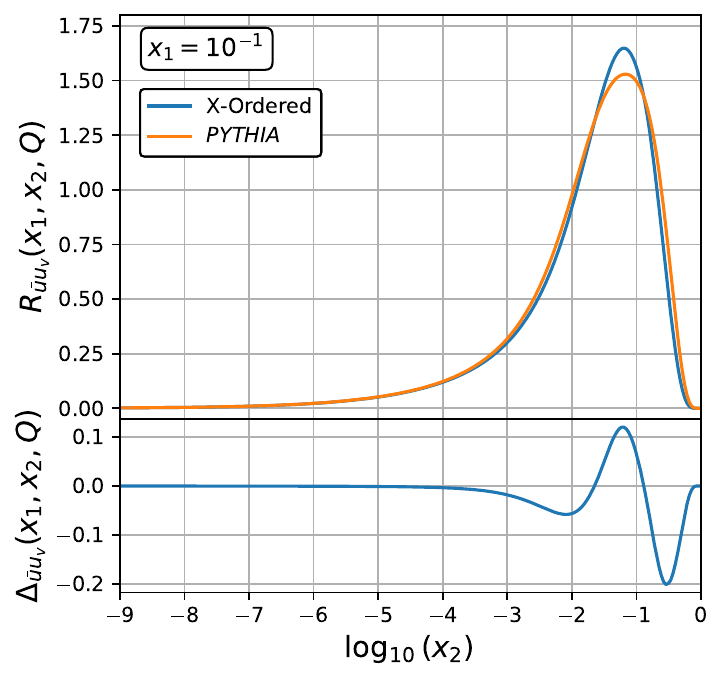}
\end{subfigure}
\caption{Behavior of  the Number Sum Rule response functions for $\bar{u}u_v$.}
\label{ubuv_rfn_plots}
\end{figure}

In Fig.~\ref{ubuv_rfn_plots} we make these plots for the $\bar{u}u_v$ sum rule. In both cases one observes a bump when $x_2 \simeq x_1$, which corresponds to the companion of the $\bar{u}$ quark - although as $x_1$ approaches $0.1$ this bump becomes more difficult to discern on top of the valence quark peak. However, what is quite noticeable is that the bumps in the X-ordered case are narrower and sharper than that for the \pythia case; in the \pythia case the companion has an extra factor of a sea quark PDF that appears to smear out the $x$ values for the companion. We thus observe that our X-ordered companion mechanism has a rather stronger preference than the \pythia one to produce the companion quark with a similar $x$ value as that of the sea quark.

\begin{figure}
\centering
\includegraphics[width=0.8\textwidth]{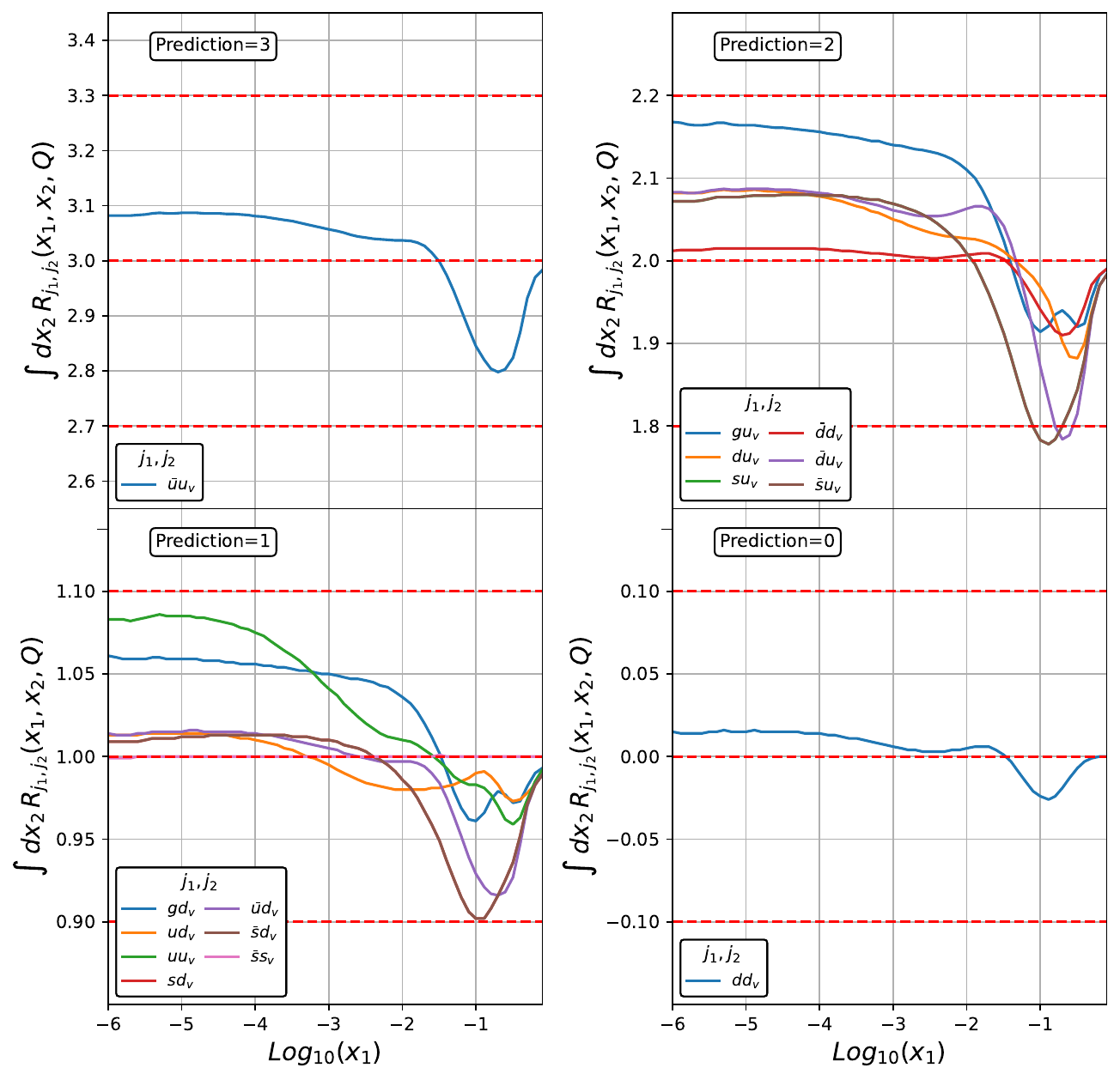}
\caption{Behavior of the Number Sum Rule integrals for all parton flavors in the X-ordered dPDF model at $Q=2$ GeV.}
\label{fig:NSR2GeV}
\end{figure}

\begin{figure}
\centering
\includegraphics[width=0.6\textwidth]{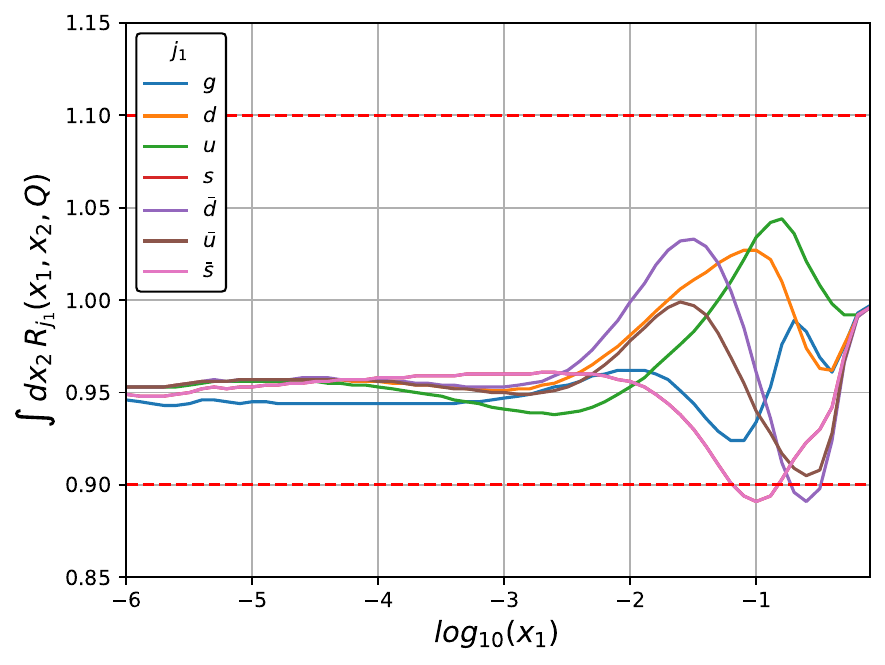}
\caption{Behavior of the Momentum Sum Rule integrals for all parton flavors in the X-ordered dPDF model at $Q=2$ GeV.}
\label{fig:MSR2GeV}
\end{figure}

The results at $Q=2$ GeV for the number and momentum sum rules are given in Figs.~\ref{fig:NSR2GeV} and \ref{fig:MSR2GeV}. We observe that once the damping in \eqref{lowxgdamp} is applied, the sum rules are almost entirely within 10\% of 1, although the curves are generally closer to the $\pm 10 \%$ boundaries than in the $Q=M_Z$ case. In fact, the $\bar{d}u_v$ and $\bar{s}u_v$ number curves as well as the $\bar{s}$ and $\bar{d}$ momentum curves do go slightly beyond the $-10\%$ threshold at $x_1 \sim 0.1$.

Why are the deviations generally larger in this case? Let us focus first on the number sum rules. At $Q = 2$ GeV the $a_v$ factor is much larger than at $Q = M_Z$ ($\sim 1.2$ rather than $\sim 1.1$ for the $u$ and $d$), meaning that, before damping, the sum rule curves at low $x_1$ are significantly larger than at $Q = M_Z$ (as we saw in Tables \ref{XordNStable} and \ref{XordNS2GeVgluontable}). At high $x_1$ there is some compensation between the overall $a_v$ factor enhancement, and a tendency for the large $x_2$ suppressive troughs in the lower panes of Fig.~\ref{ubdv_otherx1_plot} to increase, which is linked to the greater importance of the high $x$ region at smaller $Q$. We see in  Tables \ref{XordNStable} and \ref{XordNS2GeVgluontable} that before damping, the troughs at around $x_1 \sim 0.1$ have the same size for $Q = 2$ GeV and $Q= M_Z$. Thus, already before damping, the sum rules vary over a larger range at $Q = 2$ GeV than $Q=M_Z$. We have to apply a more aggressive damping to pull the low $x_1$ behavior inside the $+10\%$ bound, and this more aggressive damping also impacts the higher $x_1$ regions of the sum rule curves, pushing the trough down towards the $-10\%$ bound. 

The increased size of the troughs for some of the momentum sum rules in Fig.~\ref{fig:MSR2GeV} is likely caused by a subset of the effects discussed for the number sum rules; namely, a greater prominence of the higher $x$ region, and a more aggressive damping.
One could presumably pull the sum rule curves that stray slightly outside the $\pm 10\%$ bounds back inside the bounds by using flavour-dependent damping factors -- however we refrain from doing so here, as this would make the construction more complex, and the violations are only slightly beyond $10\%$ for a small range of $x_1$ around $x_1 \sim 0.1$. We also note that the sum rules in which the strongest violations appear at $x_1 \sim 0.1$ are the least important ones at this $x_1$ value, since at $x_1 \sim 0.1, Q = 2$ GeV, the $u,d$ and $g$ PDFs are much larger than the sea quark PDFs. We note that the sum rules with the $u,d$ and $g$ flavours as parton $1$ do not violate the $\pm 10 \%$ threshold at high $x_1$.

\begin{figure}
\centering
\includegraphics[width=0.8\textwidth]{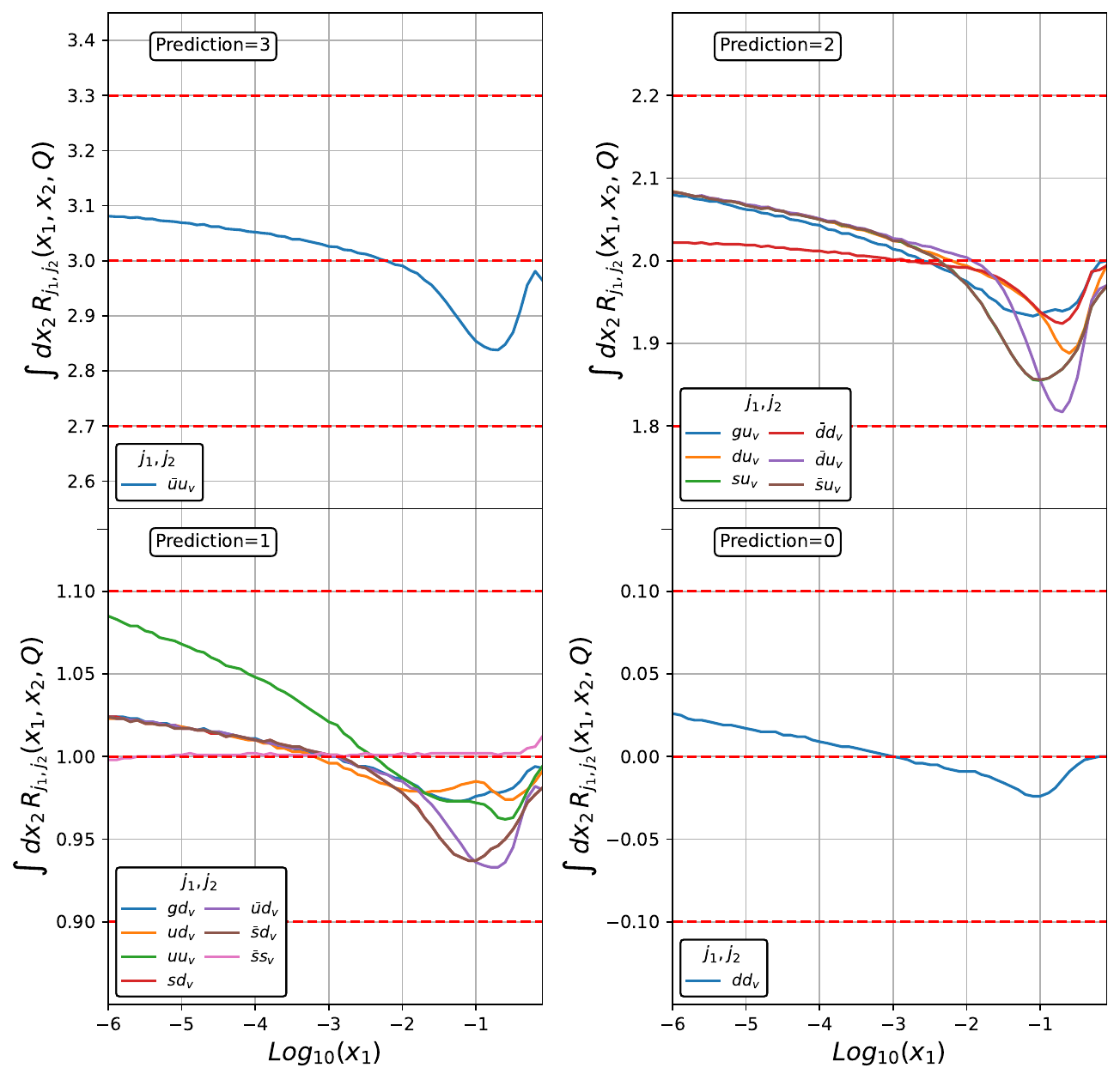}
\caption{Behavior of the Number Sum Rule integrals for dPDFs that have been obtained by evolving an X-ordered construction at $Q = 2$ GeV up to $Q=M_Z$ GeV, using inhomogeneous double DGLAP evolution.}
\label{fig:XinitNSREvals_Mz}
\end{figure}

\begin{figure}
\centering
\includegraphics[width=0.6\textwidth]{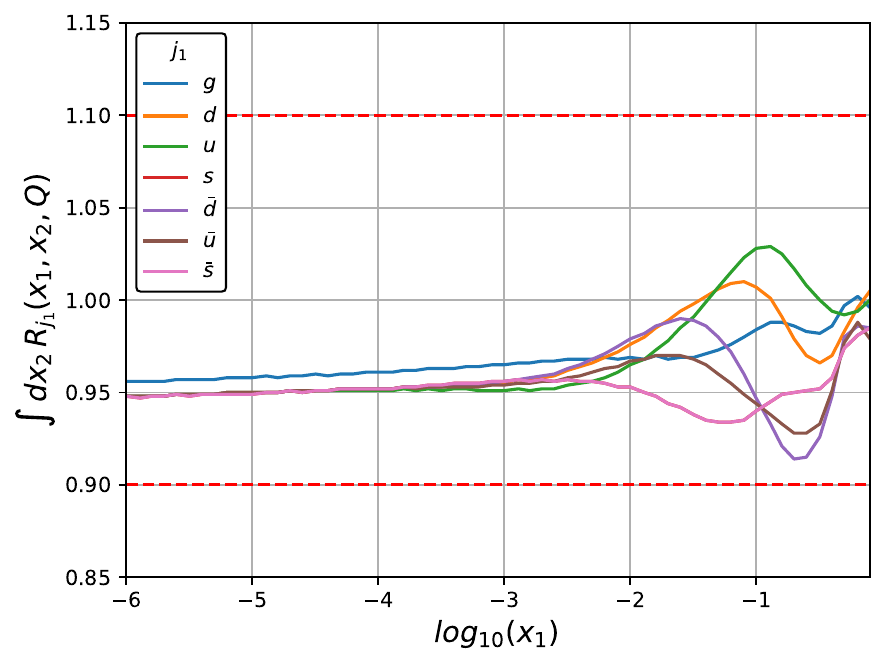}
\caption{Behavior of the Momentum Sum Rule integrals for dPDFs that have been obtained by evolving an X-ordered construction at $Q = 2$ GeV up to $Q=M_Z$ GeV, using inhomogeneous double DGLAP evolution.}
\label{fig:XinitMSREvals_Mz}
\end{figure}

Let us now take these dPDFs constructed at \mbox{$Q=2$ GeV}, evolve them up to $Q=M_Z$ using the inhomogeneous dDGLAP equation, and study how well the evolved dPDFs satisfy the sum rules.
To achieve this evolution numerically, we use the \texttt{DOVE} code that was developed in Ref.~\cite{Gaunt:2009re}. For the evolution, 240 points were used in each $x_i$ direction spanning $10^{-9} < x_i < 1$, and 60 points in the $t \equiv \log(Q^2)$ direction spanning $2 \text{ GeV} <  Q < 100 \text{ GeV}$. We take $m_c = 1.4$ GeV ($< 2$ GeV), $m_b = 4.75$ GeV, and $\alpha_s(M_Z) = 0.135$ as in the MMHT2014LO set. From the resultant grid of dPDF values we extracted the dPDFs at $Q=M_Z$ using the GS09 interpolation code \cite{Gs09web}, and we plot the sum rule curves for these dPDFs in Figures~\ref{fig:XinitNSREvals_Mz} and \ref{fig:XinitMSREvals_Mz}. For consistency, the single PDFs used in these sum rule curves have been obtained by evolving the MMHT2014LO PDFs at $Q=2$ GeV up to $M_Z$ using the same \texttt{DOVE} code; these differ at the $O(1\%)$ level from the MMHT2014LO PDFs at $Q=M_Z$ due to differences in the interpolation routine, and formally higher order differences in the treatment of the strong coupling constant $\alpha_s$.

Comparing to Figures~\ref{fig:NSR2GeV} and \ref{fig:MSR2GeV}, we see that for most number sum rule curves evolution smoothes out the curves and pulls them closer to the expected values (note that in particular, there are no longer any fluctuations outside the bounds). Note that the $\bar{s}s_v$ curve still lies perfectly at $1$ (up to some small numerical artifacts near $x_1 = 1$) due to the fact that this satisfies the sum rules perfectly at $Q=2$ GeV, and then is not fed by any other dPDF during evolution (note that the $gs_v$ distribution is zero).

\begin{figure}
\centering
\includegraphics[width=0.79\textwidth]{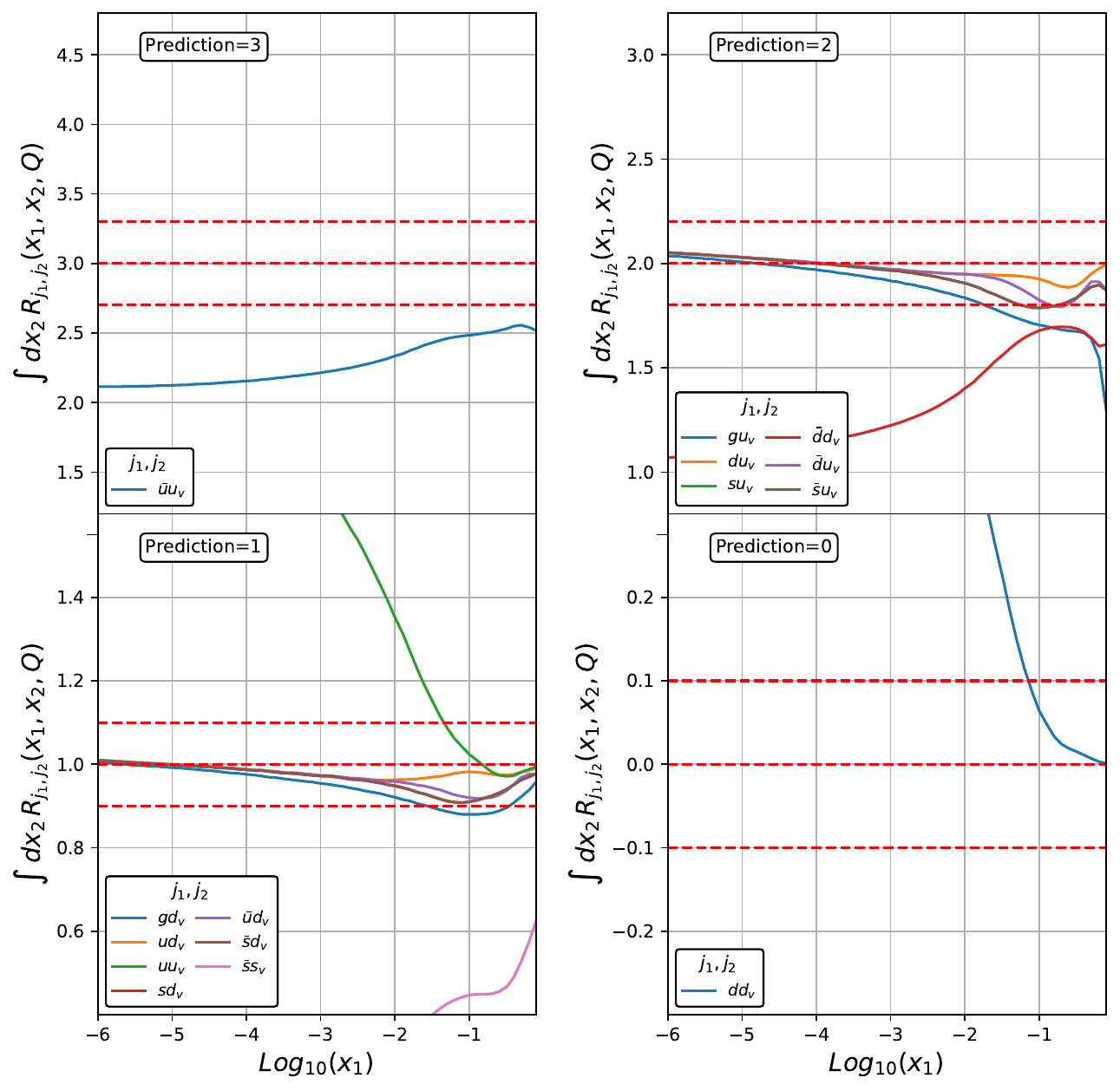}
\caption{Behavior of the Number Sum Rule integrals for dPDFs that have been obtained by evolving an X-ordered construction at $Q = 2$ GeV up to $Q=M_Z$ GeV, using homogeneous double DGLAP evolution without the ``$1 \to 2$'' term.}
\label{fig:XinitNSREvals_int_Mz}
\end{figure}

\begin{figure}
\centering
\includegraphics[width=0.59\textwidth]{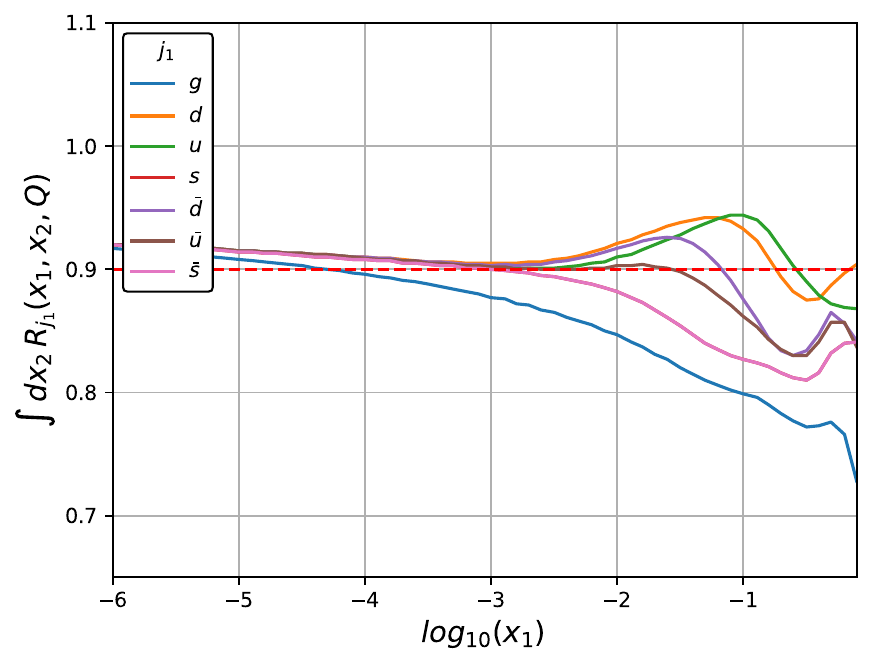}
\caption{Behavior of the Momentum Sum Rule integrals for dPDFs that have been obtained by evolving an X-ordered construction at $Q = 2$ GeV up to $Q=M_Z$ GeV, using homogeneous double DGLAP evolution without the ``$1 \to 2$'' term.}
\label{fig:XinitMSREvals_int_Mz}
\end{figure}

As an aside, we can evaluate the sum rules for dPDFs evolved with solely homogeneous evolution (\textit{i.e.} with the final ``$1\to2$'' term on the right hand side of Eq.~\eqref{eq:double_dglap} removed). 
The results are shown in Figures~\ref{fig:XinitNSREvals_int_Mz} and \ref{fig:XinitMSREvals_int_Mz}.
We see from these plots that the ``$1 \to 2$'' feed is critical to maintaining the sum rules (as has been previously mentioned in Refs.~\cite{Gaunt:2009re, Blok:2013bpa, Ceccopieri:2014ufa, Diehl:2018kgr, Diehl:2020xyg}). For the number sum rule curves, the curves which are particularly drastically affected are those rules involving a $q\bar{q}$ dPDF (\textit{e.g.} the $uu_v,dd_v, \bar{s}s_v$ sum rules), since that type of dPDF is directly fed by the gluon sPDF (whilst the other $qq$ dPDF in the sum rule is not). We note that the momentum sum rule curves are now lying at $\sim 0.9$ rather than $\sim 0.95$; the ``$1 \to 2$'' feed injects partons, and thus momentum, into the dPDFs, so if we remove this feed the momentum sum rule curves will gradually drift downwards as $Q$ increases.

%% file: Sections/TPD_Sum_Rules.tex
\subsection{Construction of the triple parton distribution functions}
\label{sec:tpdfs_results}
The generalisation of this procedure to higher multiplicity mPDFs is fairly straightforward, and we detail it here for the tPDF case. We can define $T^{\{1,2,3\}}_{j_1j_2j_3}(x_1,x_2,x_3)$ to be the modified version of the \pythia tPDF where the ordering is strictly determined by the $x$ fractions (from highest to lowest), and we replace their companion quark PDFs with ours (as defined in equation \eqref{eq:newcompanion}, albeit now including appropriate momentum squeezing factors).
Then, the appropriate smoothed version of these tPDFs can be constructed according to
\begin{align}
    T_{j_1j_2j_3}(x_1,x_2,x_3) =& \left\{1 - \tfrac{1}{2} \left[ F\left(\tfrac{x_{\rm mid}}{x_{\rm min}}\right) \! + \!F\left(\tfrac{x_{\rm max}}{x_{\rm mid}}\right) \right]  + \tfrac{1}{6} F\left(\tfrac{x_{\rm mid}}{x_{\rm min}}\right) F\left(\tfrac{x_{\rm max}}{x_{\rm mid}}\right) \right\} T^{\{1,2,3\}}_{j_1j_2j_3}(x_1,x_2,x_3)
    \nonumber \\
    &+\left\{ \tfrac{1}{2}F\left(\tfrac{x_{\rm mid}}{x_{\rm min}}\right) - \tfrac{1}{3}F\left(\tfrac{x_{\rm mid}}{x_{\rm min}}\right)F\left(\tfrac{x_{\rm max}}{x_{\rm mid}}\right)\right\} T^{\{1,3,2\}}_{j_1j_2j_3}(x_1,x_2,x_3)
    \nonumber \\
    &+\left\{ \tfrac{1}{2}F\left(\tfrac{x_{\rm max}}{x_{\rm mid}}\right) - \tfrac{1}{3}F\left(\tfrac{x_{\rm mid}}{x_{\rm min}}\right)F\left(\tfrac{x_{\rm max}}{x_{\rm mid}}\right)\right\} T^{\{2,1,3\}}_{j_1j_2j_3}(x_1,x_2,x_3)
    \nonumber \\
    &+\tfrac{1}{6}F\left(\tfrac{x_{\rm mid}}{x_{\rm min}}\right)F\left(\tfrac{x_{\rm max}}{x_{\rm mid}}\right) \Big\{ T^{\{3,2,1\}}_{j_1j_2j_3}(x_1,x_2,x_3) + T^{\{3,1,2\}}_{j_1j_2j_3}(x_1,x_2,x_3) 
    \nonumber \\
    &+ T^{\{2,3,1\}}_{j_1j_2j_3}(x_1,x_2,x_3)\Big\},
\end{align}
where $x_{\rm min} = \min(x_1,x_2,x_3), \, x_{\rm max} = \max(x_1,x_2,x_3)$, and $x_{\rm mid}$ is the remaining $x_i$ value in between $x_{\rm min}$ and $x_{\rm max}$. 
The quantity $T^{\{i,j,k\}}_{j_1j_2j_3}(x_1,x_2,x_3)$ is the ordered tPDF where the parton with the $i^{\rm th}$ largest $x$ value is first in the ordering, the $j^{\rm th}$ largest $x$ value is second, and the $k^{\rm th}$  is last. The function $F$ is the smoothing function defined in Eq.~\eqref{eq:smoothingfunc}.

We construct the tPDFs in this way so that they are smooth and satisfy the following:
\begin{align}
  T_{j_1j_2j_3}(x_1,x_2,x_3) & \xrightarrow[x_{\rm max} \gg x_{\rm mid} \gg x_{\rm min}]{} && T^{\{1,2,3\}}_{j_1j_2j_3}(x_1,x_2,x_3),
\\
    T_{j_1j_2j_3}(x_1,x_2,x_3) &\xrightarrow[x_{\rm max} \gg x_{\rm mid} = x_{\rm min}]{} && \tfrac{1}{2}\left\{T^{\{1,2,3\}}_{j_1j_2j_3}(x_1,x_2,x_3)+
T^{\{1,3,2\}}_{j_1j_2j_3}(x_1,x_2,x_3)\right\},
\\
    T_{j_1j_2j_3}(x_1,x_2,x_3) &\xrightarrow[x_{\rm max} = x_{\rm mid} \gg x_{\rm min}]{} && \tfrac{1}{2}\left\{T^{\{1,2,3\}}_{j_1j_2j_3}(x_1,x_2,x_3)+
T^{\{2,1,3\}}_{j_1j_2j_3}(x_1,x_2,x_3)\right\},
\\
    T_{j_1j_2j_3}(x_1,x_2,x_3) & \xrightarrow[x_{\rm max} = x_{\rm mid} = x_{\rm min}]{} && \tfrac{1}{6}\Big\{T^{\{1,2,3\}}_{j_1j_2j_3}(x_1,x_2,x_3)+
T^{\{1,3,2\}}_{j_1j_2j_3}(x_1,x_2,x_3)
\nonumber \\
&&& +T^{\{2,1,3\}}_{j_1j_2j_3}(x_1,x_2,x_3) 
 + T^{\{3,2,1\}}_{j_1j_2j_3}(x_1,x_2,x_3) 
\nonumber \\
&&& + T^{\{3,1,2\}}_{j_1j_2j_3}(x_1,x_2,x_3) + T^{\{2,3,1\}}_{j_1j_2j_3}(x_1,x_2,x_3)
    \Big\}.
\end{align}
That is, when a parton has an $x$ fraction that is very different from the others, it participates in the X-ordering procedure, but if we have two (or more) partons with the same $x$ fractions, we recover the naive symmetrisation procedure for those partons.

The last step to achieve the full X-ordered tPDFs is to apply the damping factor pairwise. That is, we consider the three possible pairings of partons in the tPDF, and apply the dPDF damping factor (introduced in equation \eqref{lowxgdamp}) to each pair. For the parts of the tPDF that involve a companion quark, we choose to not introduce any damping factors.

A numerical study of the sum rules for the tPDF case is a more involved exercise than the study for the dPDF case, due to the fact that we now have two ``spectator partons'' with two possible $x$ values and flavours. Thus, for simplicity we shall only make a tPDF study at $Q=M_Z$ -- we expect the extent to which the X-ordered tPDFs satisfy the sum rules at lower scales to be slightly worse, as we saw for the dPDFs. We construct the tPDFs at $Q = M_Z$ using the procedure just described, and then evaluate the tPDF sum rule integrals for a grid of points in the $x_1,x_2$ plane. We choose the points on the grid to have $x_i$ values in the set $\left\{10^{-6},10^{-4},10^{-3},0.1,0.2,0.4,0.8\right\}$ (and discard any points with $x_1 + x_2 \ge 1$). This gives us $44$ points in the $x_1,x_2$ plane. For each point we compute the ``percentage deviation'' from expectation, where this is defined by an analogous procedure to the one given at the start of Section~\ref{sec:Xord_dPDF_construction} for the dPDFs. The dPDFs we use in this computation are our final X-ordered ones constructed in Section~\ref{sec:Xord_dPDF_construction}.

\begin{figure}
    \centering
    \captionsetup{width=0.8\textwidth}
    \includegraphics[width=\linewidth]{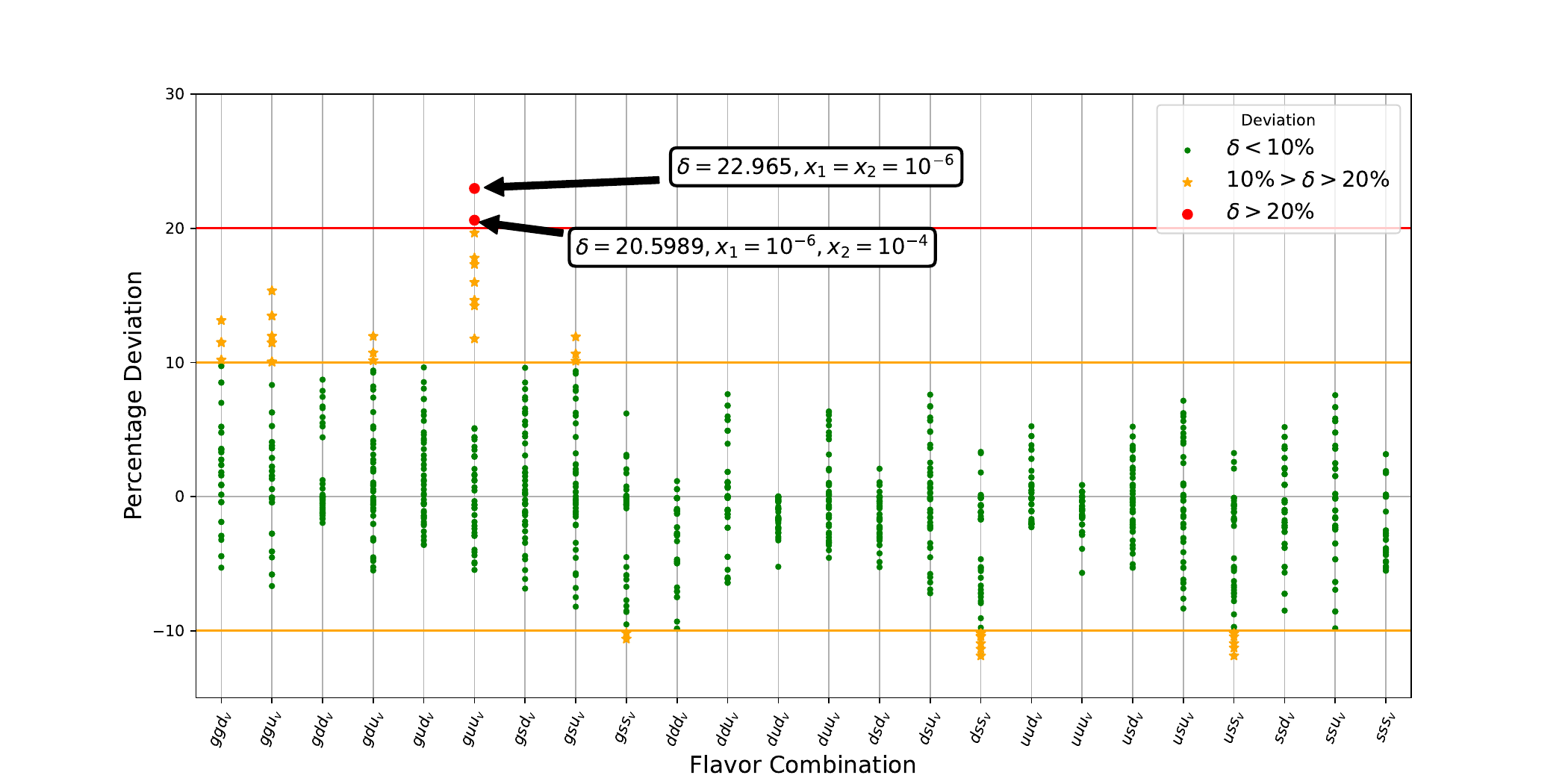}
    \caption{A survey of the $j_1j_2j_{3_v}$ tPDF Number Sum rules.}
    \label{jjj_rules}
\end{figure}

\begin{figure}
    \centering
    \captionsetup{width=0.8\textwidth}
    \includegraphics[width=\linewidth]{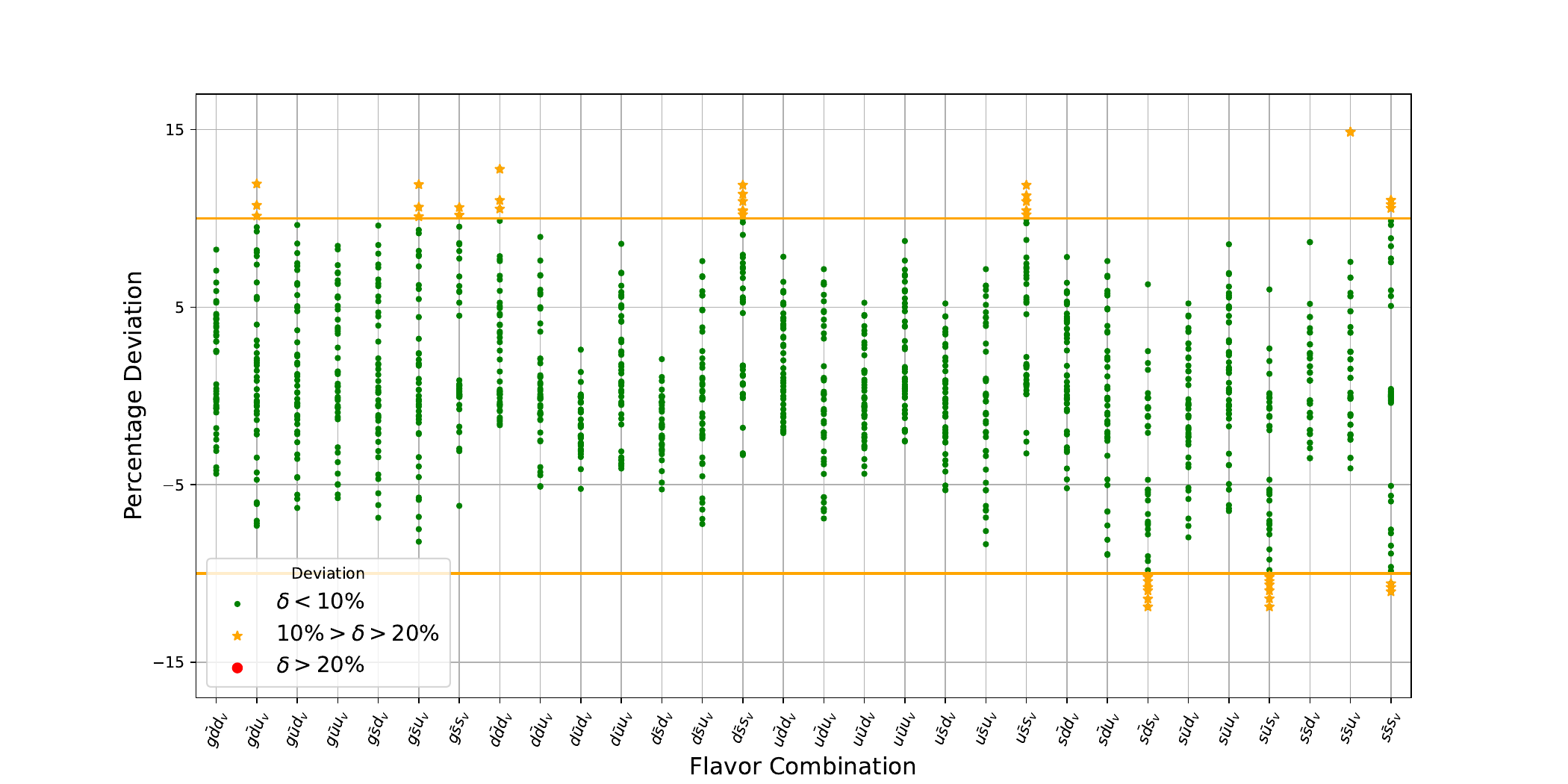}
    \caption{A survey of the $j_1\bar{j}_2j_{3_v}$ tPDF Number Sum rules.}
    \label{jjbj_rules}
\end{figure}

\begin{figure}
    \centering
    \captionsetup{width=0.8\textwidth}
    \includegraphics[width=\linewidth]{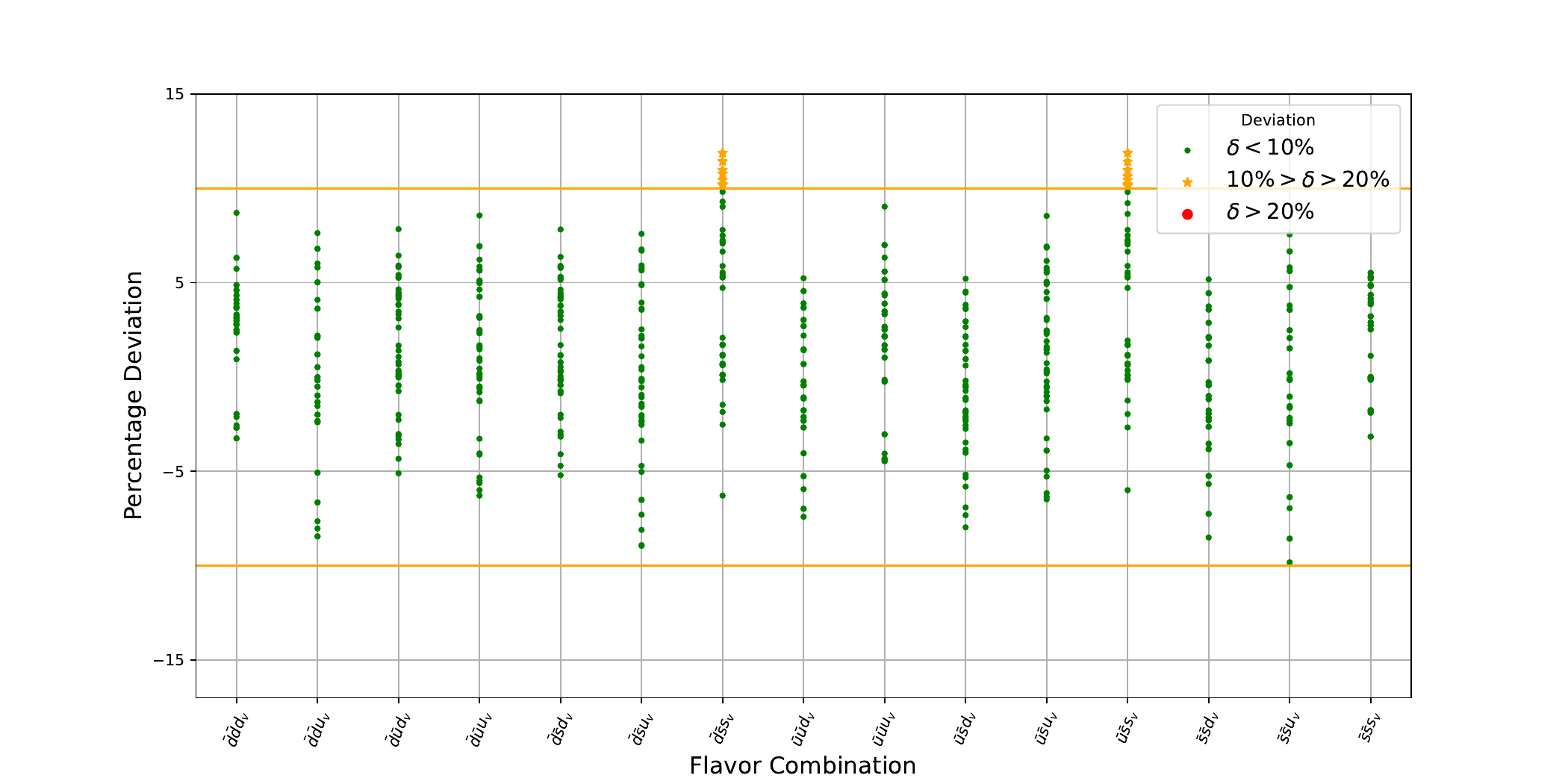}
    \caption{A survey of the $\bar{j}_1\bar{j}_2j_{3_v}$ tPDF Number Sum rules.}
    \label{jbjbj_rules}
\end{figure}

We find that for $Q=M_Z$ GeV all the momentum sum rules are within $10\%$ of expectation, but there are some number sum rules that violate that threshold. The results are given for the number sum rules in Figures \ref{jjj_rules}, \ref{jjbj_rules} and \ref{jbjbj_rules}. For each sum rule (listed along the $x$-axis), we plot as a vertical scatter the percentage deviation for the 44 $\{x_1,x_2\}$ points; this gives an overall picture of how well the sum rules are satisfied across the phase space. For clarity, the $\{x_1,x_2\}$ values of the points are labeled only for those whose percentage deviations exceed $20\%$. We see that the vast majority of points do fall within the $10\%$ envelope, and for all but one of the sum rules, the other points only lie slightly outside this envelope. The exception to this is the $guu_v$ number sum rule where we see $>20\%$ deviations at the smallest $\{x_1,x_2\}$ values. In Ref.~\cite{Fedkevych:2022myf} we tested the $uuu_v$ and $u\bar{u}u_v$ sum rules for the naive symmetrised tPDFs and found some $> 10\%$ violations even when restricting $x_2$ to $10^{-4}$ -- looking at Figs.~\ref{jjj_rules} and \ref{jjbj_rules} we see that these sum rules are now within $10\%$ over the whole grid of points tested.

\begin{figure}
\centering
\captionsetup{width=0.8\textwidth}
\includegraphics[width=0.8\textwidth]{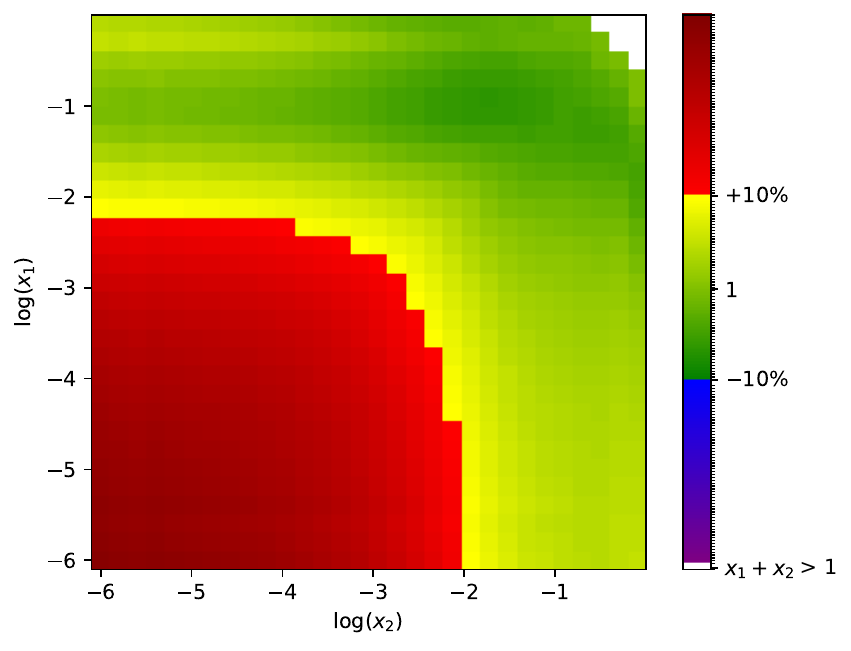}
\caption{Heat map showing the deviation from expectation of the $guu_v$ tPDF number sum rule.}
\label{tpdf_heatmaps}
\end{figure}

Since the $guu_v$ case represents something of an outlier, it is interesting to study this in more detail. In Figure~\ref{tpdf_heatmaps} we show the integral of the response function for  the $guu_v$ sum rule over the $x_1,x_2$ plane, displaying the deviations of this sum rule from the expected value of $1$. We observe that the $> 10\%$ deviations are concentrated in the region where both $x_1$ and $x_2$ are small. 

We can explain why the integral of the response function is so large for small $x_1$ and $x_2$ using similar reasoning as was used to explain why the $\bar{d}u_v$ dPDF response functon integral lies above $1$ in the discussion of Section~\ref{sec:sum_rule_assesment_dPDFs}. The important region leading to the enhancement will be $x_1,x_2 \ll x_3 \ll 1$, where the momentum squeezing effects are small. In this region our prescription is the ``wrong way round'', and in this tPDF case we now get an increased enhancement from two factors of $a_v$. On the other hand, since $k_g(M_Z)$ is essentially negligible, for $guu_v$ we effectively only have a suppression from one damping factor that involves $x_2$ and $x_3$ (note that since it involves the relatively large $x_3$, the suppression is milder than one would obtain from a damping factor which involved the small $x_1$ and $x_2$, with the latter being present for \textit{e.g.} $udu_v$). The final issue is that the $guu_v$ case has a companion quark contribution, which at low $x_1,x_2$ acts to subtract off $1$ from the response function integral -- at small $x_1,x_2$, this does so rather accurately. The remainder of the tPDF should give $2$, but due to the imperfections already discussed, differs from $2$ by some percentage. Even if these relative deviations were similar to those encountered in other cases (like $sdu_v$), the near-perfect subtraction of $1$ by the companion quark mechanism will then lead to a relative deviation from $1$ that is double the size (just because the same deviation is being compared to $1$ rather than $2$). The same sort of effect is responsible for the $uu_v$ dPDF sum rule curve having a larger positive percentage deviation than the other number sum rules at small $x_1$ in Figure \ref{XordNSREvals}. For the $guu_v$ sum rule, we thus have a ``perfect storm'' where several effects stack together to produce a particularly large positive deviation at low $x_1,x_2$ -- no other parton combination suffers from all of these effects. In principle, one could consider making some specific additional modifications to the $guu$ and $gu\bar{u}$ tPDFs to improve the extent to which the $guu_v$ sum rules are satisfied, although this is somewhat contrary to our philosophy of making simple, broad modifications to the mPDFs that are easily generalized to higher multiplicities, and also will likely degrade some of the momentum sum rules. Thus, we do not pursue it here.

%% file: Sections/PhenoDraft_new.tex
In this section we perform a simple study of the effects of the X-ordered prescription on kinematic distributions for DPS processes. The sample DPS processes we study involve the production of two electroweak bosons; specifically we choose to look at same-sign $WW$ production and double $Z$-boson production. Same-sign $WW$ production is a well-known channel to study DPS (see \textit{e.g.}~Refs.~\cite{Kulesza:1999zh, Gaunt:2010pi, Cotogno:2018mfv, Cotogno:2020iio}) -- the corresponding SPS process is suppressed by multiple coupling constants, and produces additional jets that can be used to discriminate it from the DPS signal. DPS in same-sign $WW$ production has been measured by both CMS \cite{CMS:2019jcb, CMS:2022pio} and ATLAS \cite{ATLAS:2025bcb}. Observing DPS in $ZZ$ production would be challenging due to the low DPS rate and significant SPS backgrounds -- we include this process to have a sample DPS process that is sensitive to the sea pair dPDF/companion quark PDF (note that same-sign $WW$ does not receive contributions at leading order from the $q\bar{q}$ dPDFs).

We choose to study an observable that is particularly sensitive to longitudinal correlations, such as the ones associated with number and momentum conservation. This is the rapidity asymmetry
\begin{equation}
    \mathcal{A}=\frac{\sigma(Y_1\times Y_2<0)-\sigma(Y_1\times Y_2>0)}{\sigma(Y_1\times Y_2<0)+\sigma(Y_1\times Y_2>0)},
    \label{two_particle_asymm}
\end{equation}
where $Y_1$ and $Y_2$ are the rapidities of the two bosons. If there are no correlations between the partons, $\mathcal{A}=0$, and any $\mathcal{A} \neq 0$ indicates the presence of inter-parton correlations. We study this quantity as a function of a minimum absolute rapidity cut on the bosons, $|Y_1|,|Y_2|\geq y^{\rm min}$. As $y^{\rm min}$ increases we cut out the central region of the detector and become more sensitive to dPDFs closer to the kinematic boundary $x_1 + x_2 = 1$ (either two high $x$ values in the same hemisphere contribution or one high $x$ and one small $x$ in the opposite hemisphere contribution). Close to this kinematic boundary one expects correlations to have a strong impact, such that we generically expect $\mathcal{A}$ to increase with $y^{\rm min}$.

In the same-sign $WW$ case, one cannot reconstruct this bosonic $\mathcal{A}$ in the leptonic decay channel, due to the fact that the produced neutrinos escape the detector. In this case we also consider a leptonic version of Eq.~\eqref{two_particle_asymm}, constructed from the lepton pseudorapidities \cite{Gaunt:2010pi}. One generically expects such asymmetries to be lower than the bosonic asymmetries as the decay process ``smears out'' the rapidities. If there are no correlations in the dPDF, then this asymmetry will also be $0$ of course.

The DPS cross sections are computed according to Eq.~\eqref{eq:dPDF_Scatter_noy} with $n_f = 5$. For the partonic cross sections we use the standard leading order $W^+, W^-$ or $Z$ cross sections as appropriate. For the electroweak parameters we take the following:
\begin{equation}
    \begin{aligned}
        M_Z=91.188\text{ GeV}, &\quad \Gamma_Z=2.495\text{ GeV}, \\
        M_W=80.385\text{ GeV},&\quad \Gamma_W=2.085\text{ GeV},\\
        \alpha_{EM}=7.806358\times 10^{-3},&\quad \sin^2{\theta_W}=0.2312,\\
       M_{CKM}=\begin{bmatrix}
            |V_{ud}| & |V_{us}| & |V_{ub}|\\
            |V_{cd}| & |V_{cs}| & |V_{cb}|\\
            |V_{td}| & |V_{ts}| & |V_{tb}|
        \end{bmatrix}=&
        \begin{bmatrix}
            0.097383 & 0.2243 & 0.00382\\
            0.221 & 0.975 & 0.0408\\
            0 & 0 & 0
        \end{bmatrix},
    \label{SSW_params}
    \end{aligned}
\end{equation}
where we set the last row of the CKM matrix to zero, as we do not have any top partons in our simulations.

\subsection{Same-Sign $W$ Boson Production}

We first compute the asymmetry in same-sign $WW$ obtained from 
X-ordered dPDFs directly constructed at $Q=M_W$. We will compare this with the asymmetry obtained using the \pythia dPDFs, and also with the asymmetry obtained using the GS09 dPDF set \cite{Gaunt:2009re}. The GS09 dPDFs represent a different approach to constructing dPDFs that satisfy the sum rules. At $Q_0 = 1$ GeV dPDFs that approximately satisfy the sum rules are constructed by modifying a product of single PDFs; there are valence and ``companion quark'' adjustments that are similar to what is done for our X-ordered dPDFs, but instead of the X-ordering procedure and damping factor the ansatz is multiplied by a phase space factor that suppresses large $x_1+x_2$ (see Eqs.~(3.14) and (3.15) in Ref.~\cite{Gaunt:2009re}). These inputs are then evolved to higher scales using the inhomogenous dDGLAP equation, similar to what we did in Section~\ref{sec:sum_rule_assesment_dPDFs}. For the GS09 set, the single PDFs used in the construction of the dPDFs was the leading order MSTW 2008 set \cite{Martin:2009iq}.

\begin{figure}
\centering
\subfigure[$W^+$ Asymmetry Ratios]{%
\label{W+:A}
\includegraphics[width=0.5\textwidth]{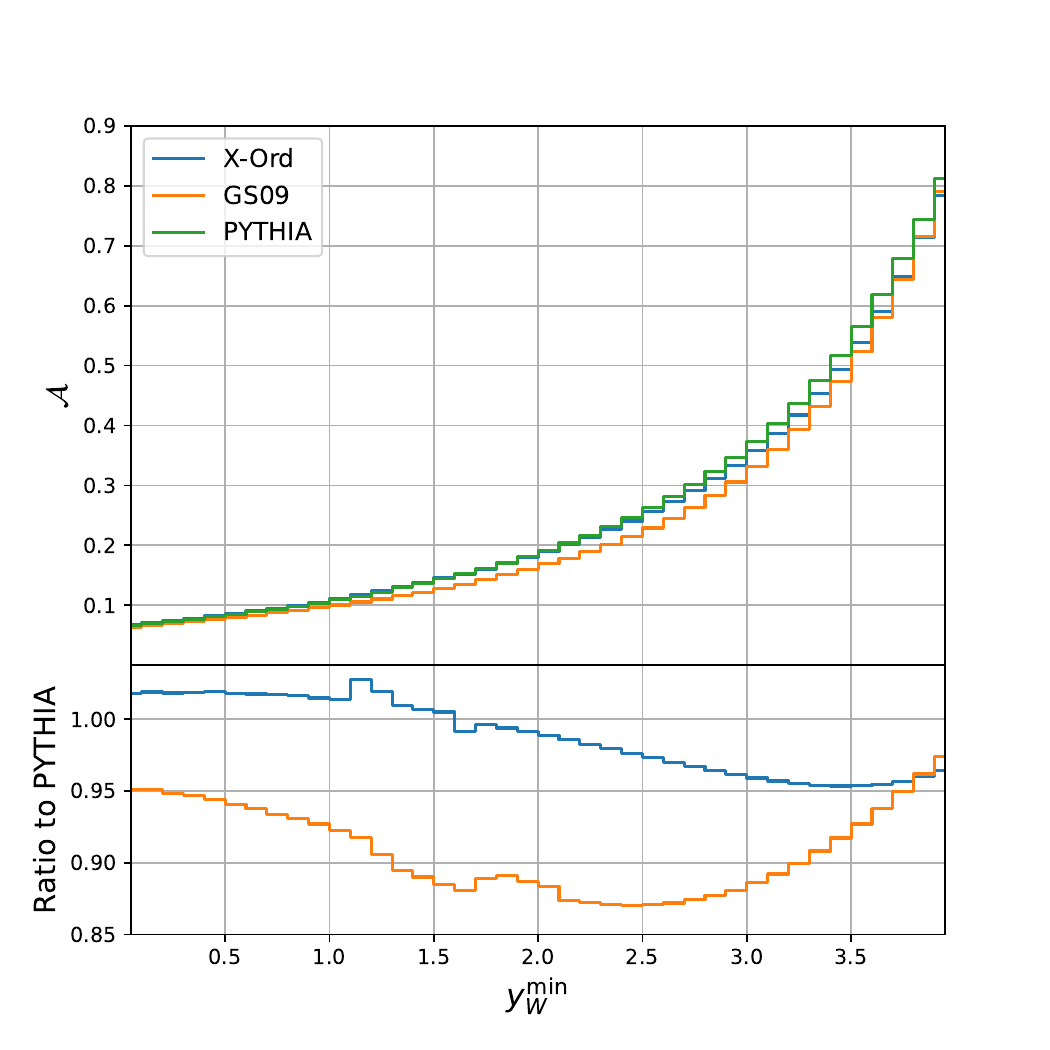}
}
\hspace{-9mm}
\subfigure[$W^-$ Asymmetries]{%
\label{W-:A}
\includegraphics[width=0.5\textwidth]{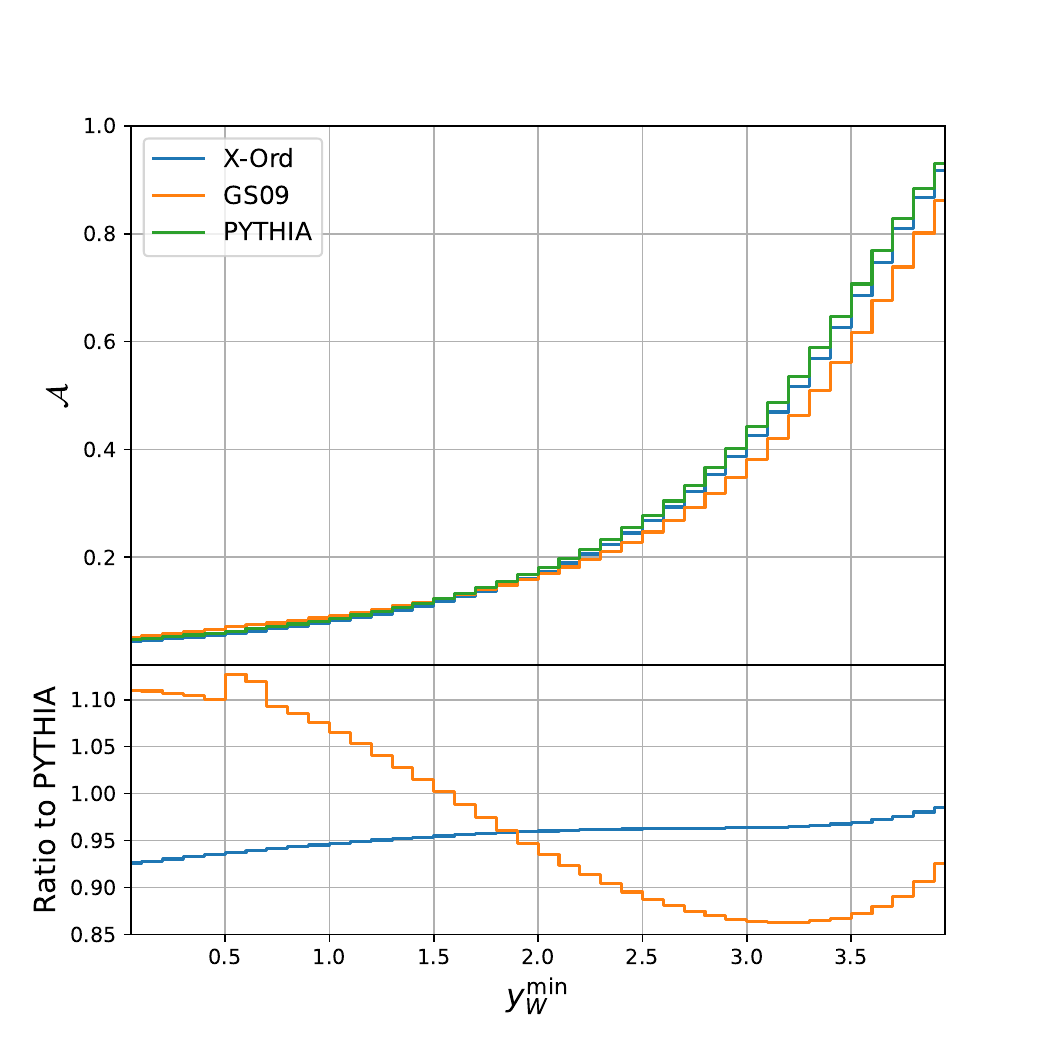}
}%
\caption{Boson asymmetries in same-sign $WW$ production, produced using \pythia, GS09, and the X-ordered dPDFs constructed at $Q=M_W$.}
\label{SSW:A}
\end{figure}

\begin{figure}
\centering
\subfigure[$W^+$ Asymmetries]{%
\label{W+:Mw_GeV_lep}
\includegraphics[width=0.5\textwidth]{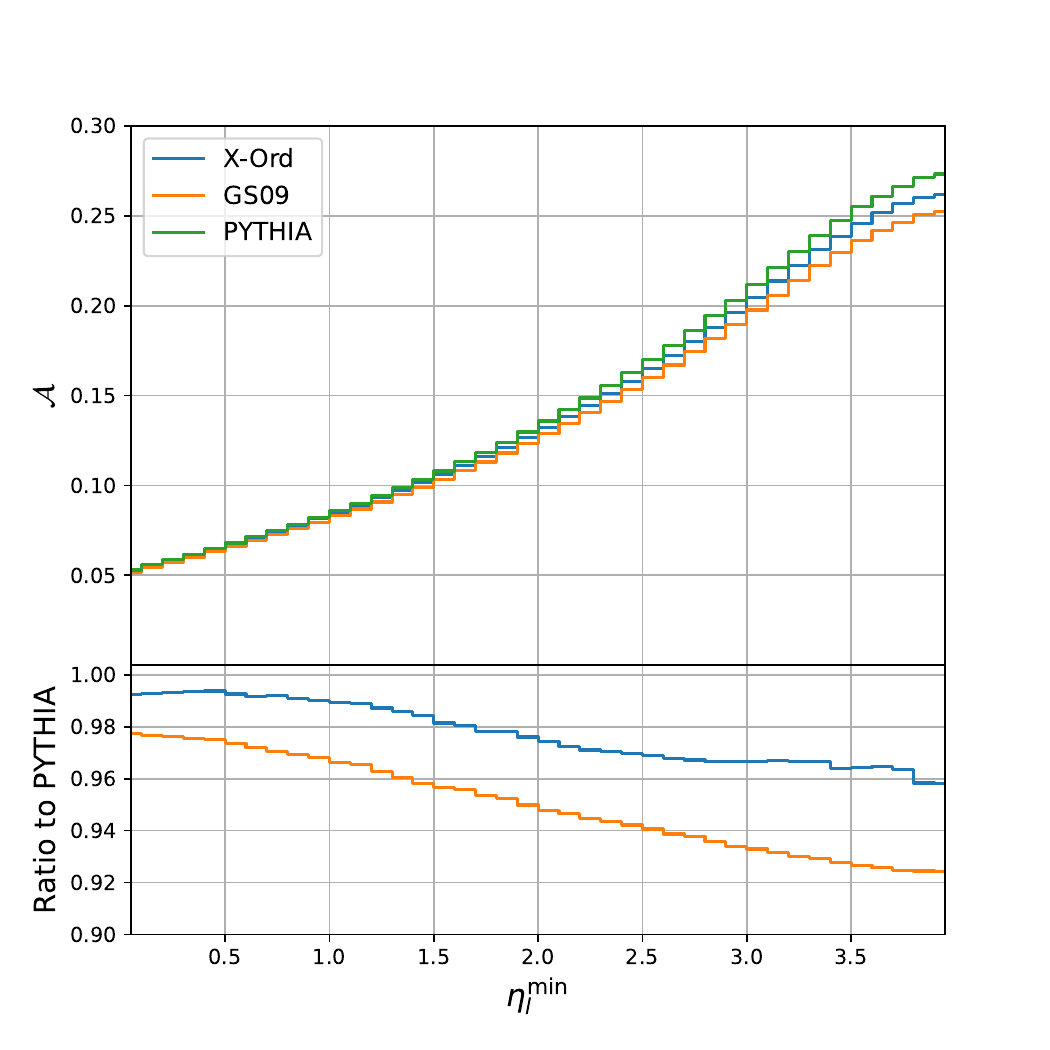}
}
\hspace{-9mm}
\subfigure[$W^-$ Asymmetries]{%
\label{W-:Mw_GeV_lep}
\includegraphics[width=0.5\textwidth]{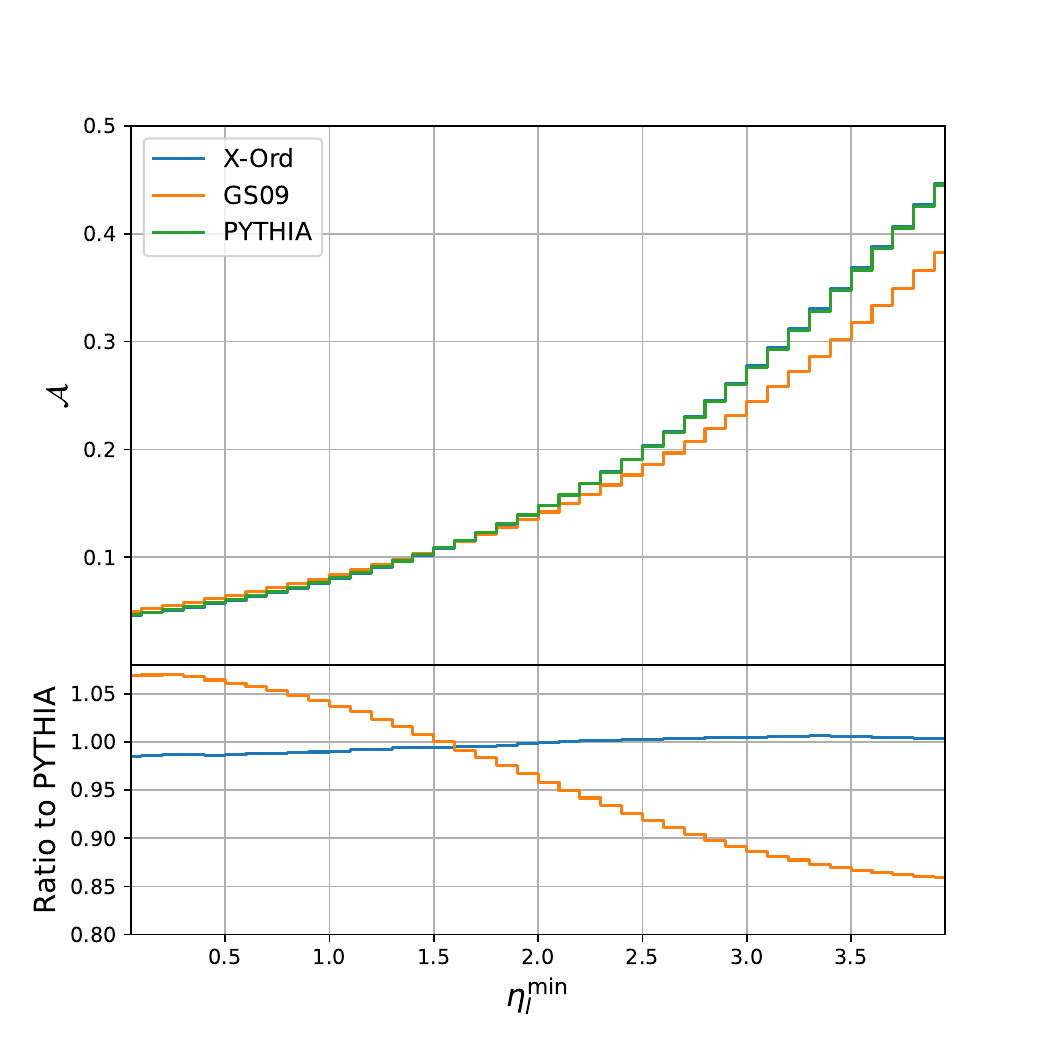}
}%
\caption{Lepton pseudorapidity asymmetries for same-sign $WW$ production, produced using \pythia, GS09, and the X-ordered dPDFs constructed at $Q=M_W$.}
\label{SSW:Mw_GeV_Lep}
\end{figure}

In Fig.~\ref{SSW:A} we plot the bosonic asymmetries, whilst the leptonic asymmetries are given in Fig. \ref{SSW:Mw_GeV_Lep}. We note that for both the bosonic and leptonic asymmetries, the results from the X-ordered dPDFs and \pythia  are in fact extremely close. We only see a small difference between the two for the $W^+$ case at large $y^{\rm min}_W/\eta_l^{\rm min}$, which can be traced to the $X$-ordering procedure. Thus, the modifications to the \pythia procedure at $Q=M_W$ to produce the X-ordered dPDFs (in particular the X-ordering procedure and the fairly weak damping modifications, since the companion mechanism is not directly probed) translate only into a minimal effect on the same-sign $WW$ asymmetries. The GS09 dPDFs yield generally lower asymmetry values than the \pythia/X-ordered results (apart from the $W^-$ case at small $y^{\rm min}_W/\eta_l^{\rm min}$) -- this was observed previously in Ref.~\cite{Cabouat:2019gtm}.

\begin{figure}
\centering
\subfigure[$W^+$ Asymmetries]{%
\label{W+:Evo_Mw_GeV}
\includegraphics[width=0.5\textwidth]{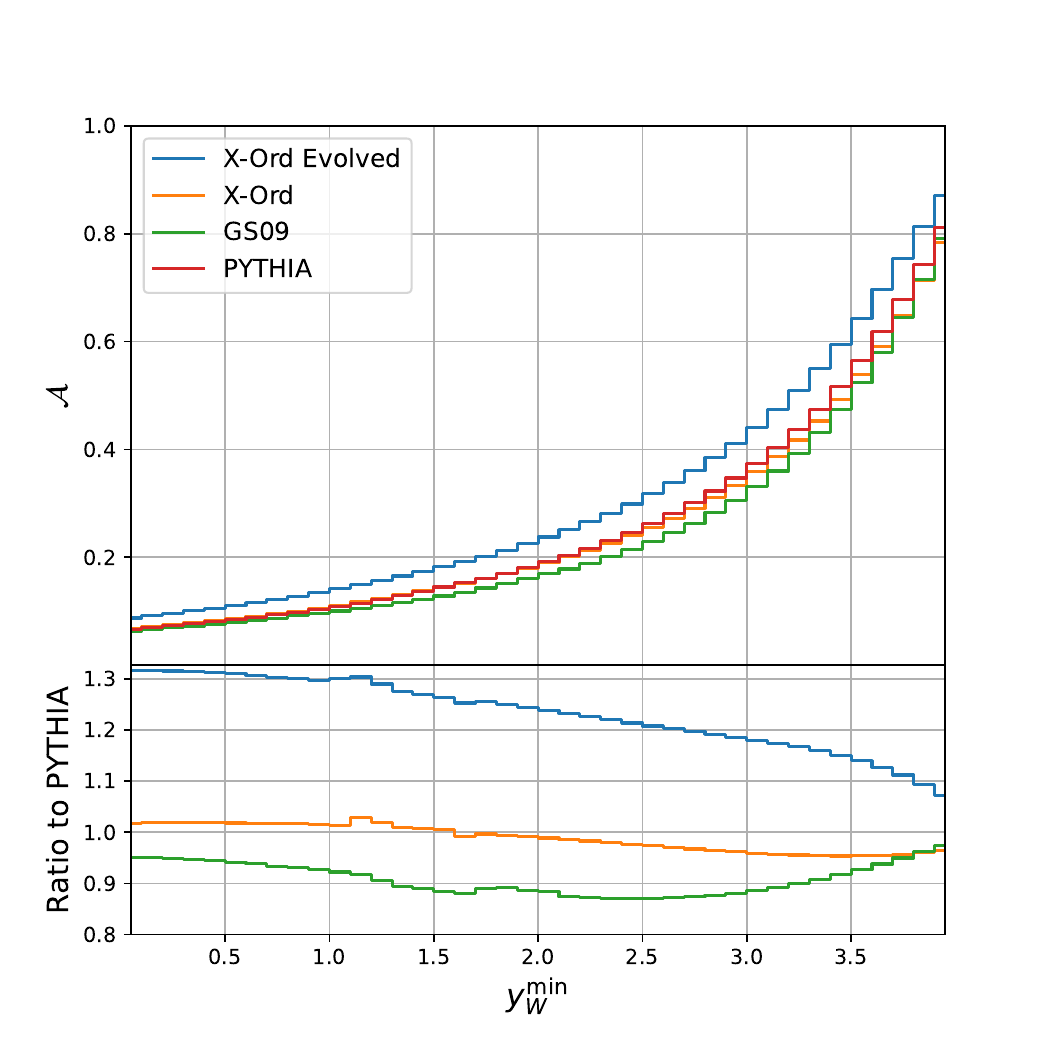}
}
\hspace{-9mm}
\subfigure[$W^-$ Asymmetries]{%
\label{W-:Evo_Mw_GeV}
\includegraphics[width=0.5\textwidth]{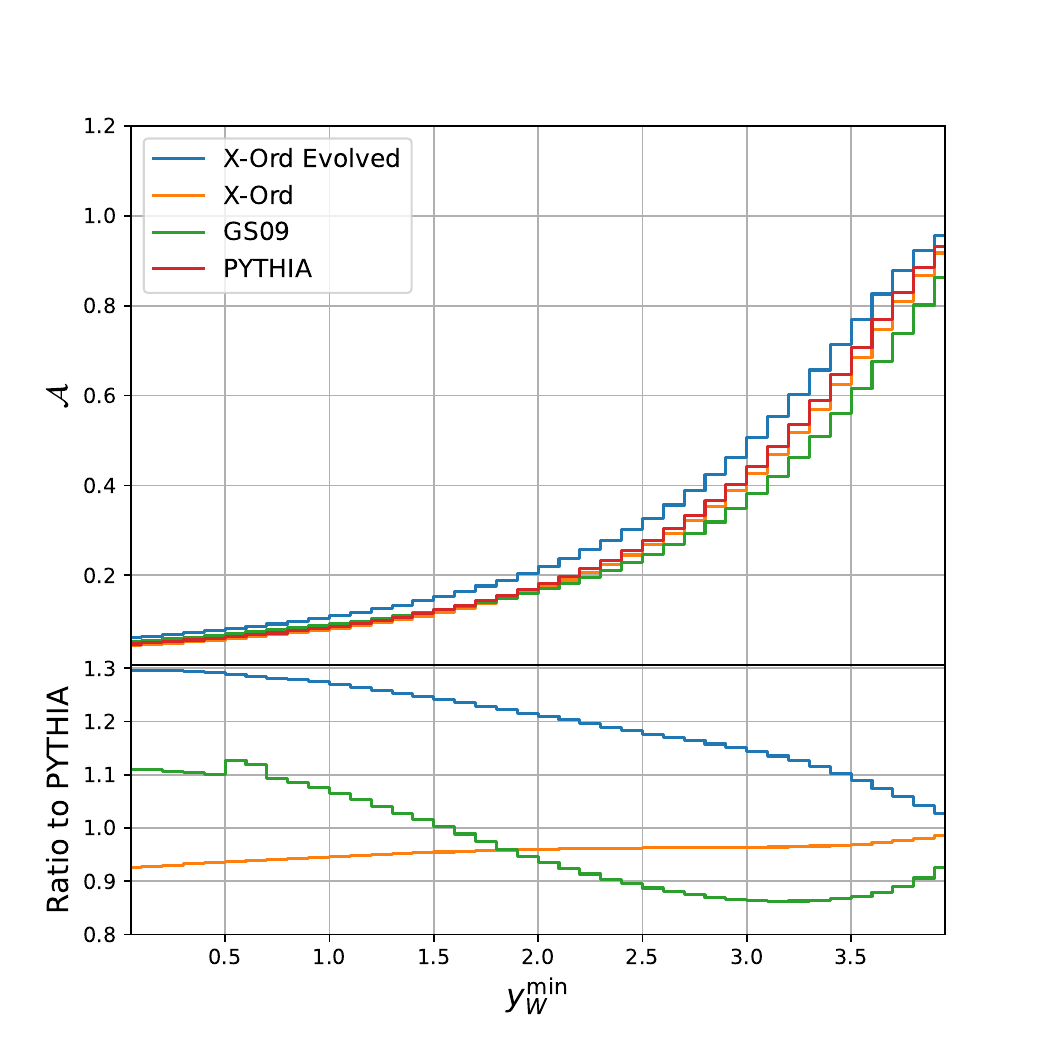}
}%
\caption{The same as in Fig. \ref{SSW:A}, but now including the bosonic asymmetry produced from dPDFs that have been constructed at $Q=2$ GeV according to the ``X-Ordering'' prescription and then evolved to $Q=M_W$ via inhomogeneous dDGLAP.}
\label{SSW:Evo_Mw_GeV}
\end{figure}

\begin{figure}
\centering
\subfigure[$W^+$ Asymmetries]{%
\label{W+:Evo_Mw_GeV_lep}
\includegraphics[width=0.5\textwidth]{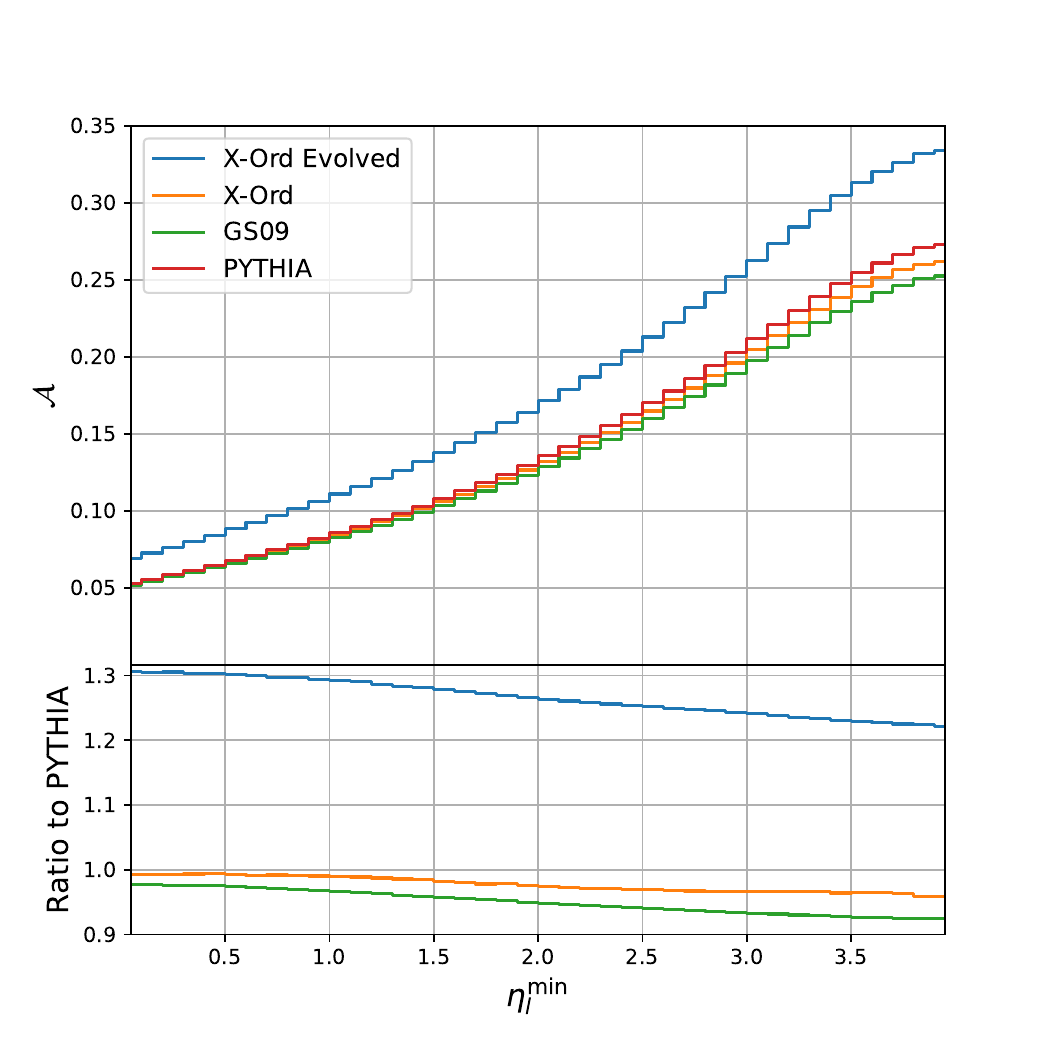}
}
\hspace{-9mm}
\subfigure[$W^-$ Asymmetries]{%
\label{W-:Evo_Mw_GeV_lep}
\includegraphics[width=0.5\textwidth]{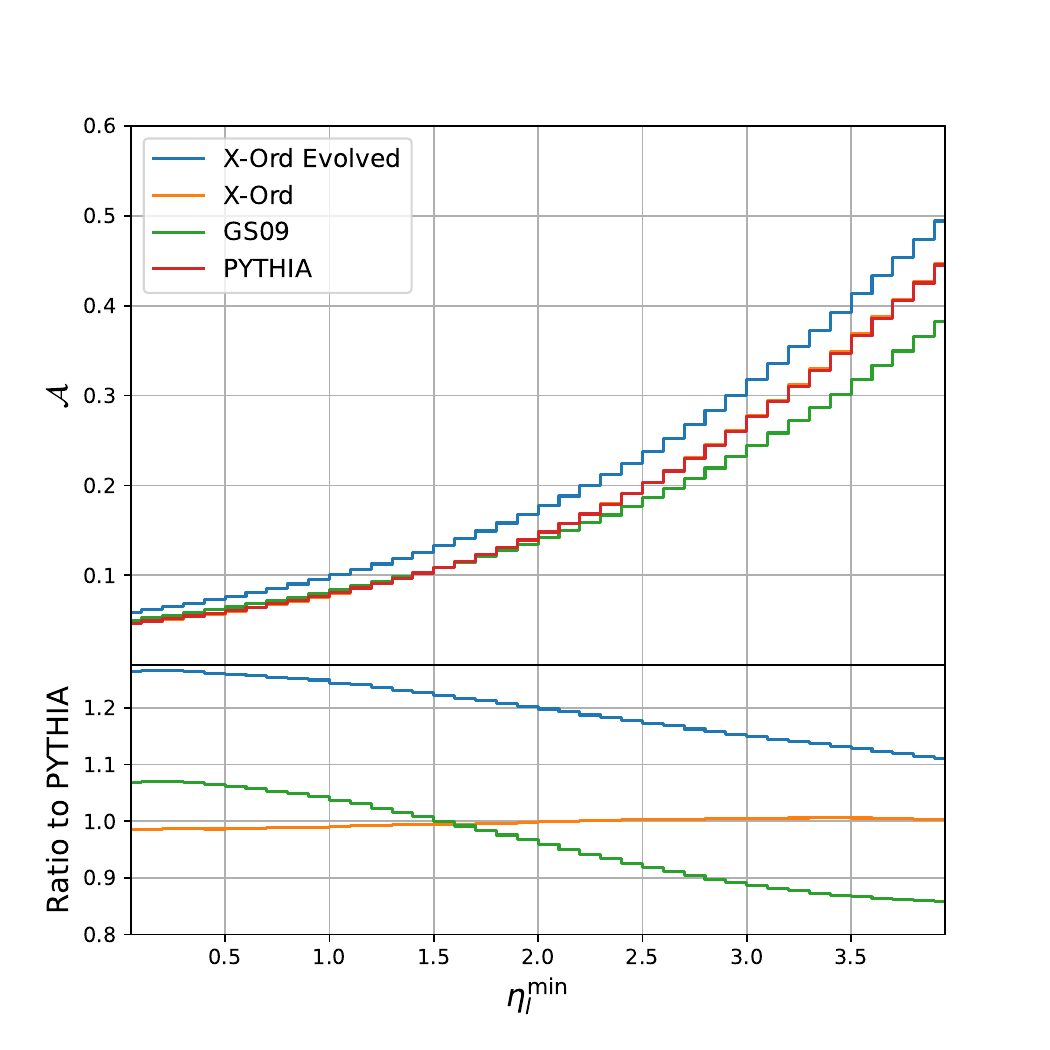}
}%
\caption{The same as in Fig.~\ref{SSW:Mw_GeV_Lep}, but now including the leptonic asymmetry produced from dPDFs that have been constructed at $Q=2$ GeV according to the X-ordering prescription and then evolved to $Q=M_W$ via inhomogeneous dDGLAP evolution.}
\label{SSW:Evo_Mw_GeV_lep}
\end{figure}

Now let us additionally compare to results obtained with X-ordered dPDFs that have been constructed at $Q=2$ GeV, and then evolved up to $Q=M_W$ using the inhomogenous dDGLAP equation (so here we are using the X-ordering procedure to construct inputs for pQCD evolution, albeit one that does not take account of the transverse correlations in the $1 \to 2$ splitting contribution). The asymmetry curves are denoted by ``X-Ord Evolved'' in Fig.~\ref{SSW:Evo_Mw_GeV} (for the bosonic asymmetry) and Fig.~\ref{SSW:Evo_Mw_GeV_lep} (for the leptonic asymmetry). In this case we see a substantial increase of the asymmetry versus the curves from \pythia or the X-ordered dPDFs constructed at $Q=M_W$. 

\begin{figure}
\centering
\subfigure[$W^+$ Asymmetries]{%
\label{W+:2GeV}
\includegraphics[width=0.5\textwidth]{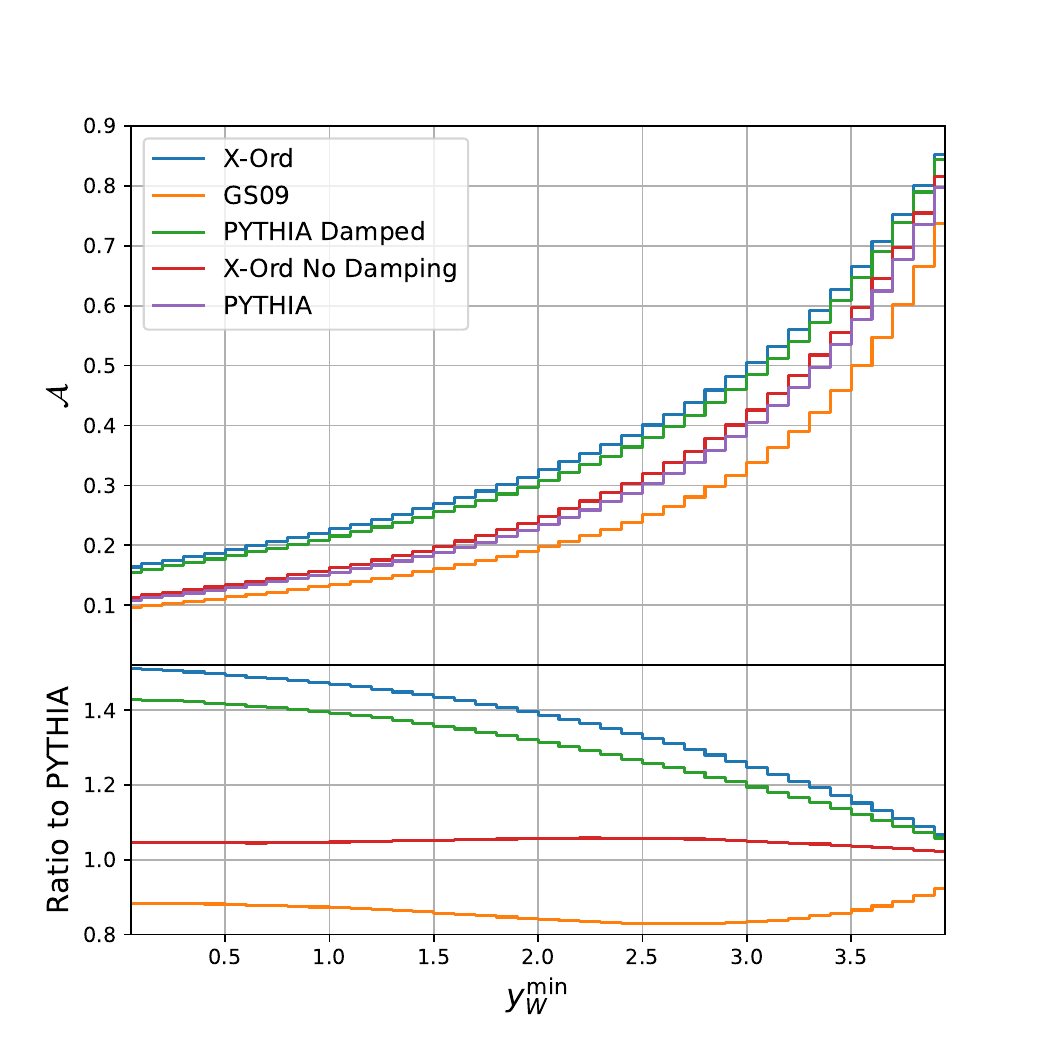}
}
\hspace{-9mm}
\subfigure[$W^-$ Asymmetries]{%
\label{W-:2GeV}
\includegraphics[width=0.5\textwidth]{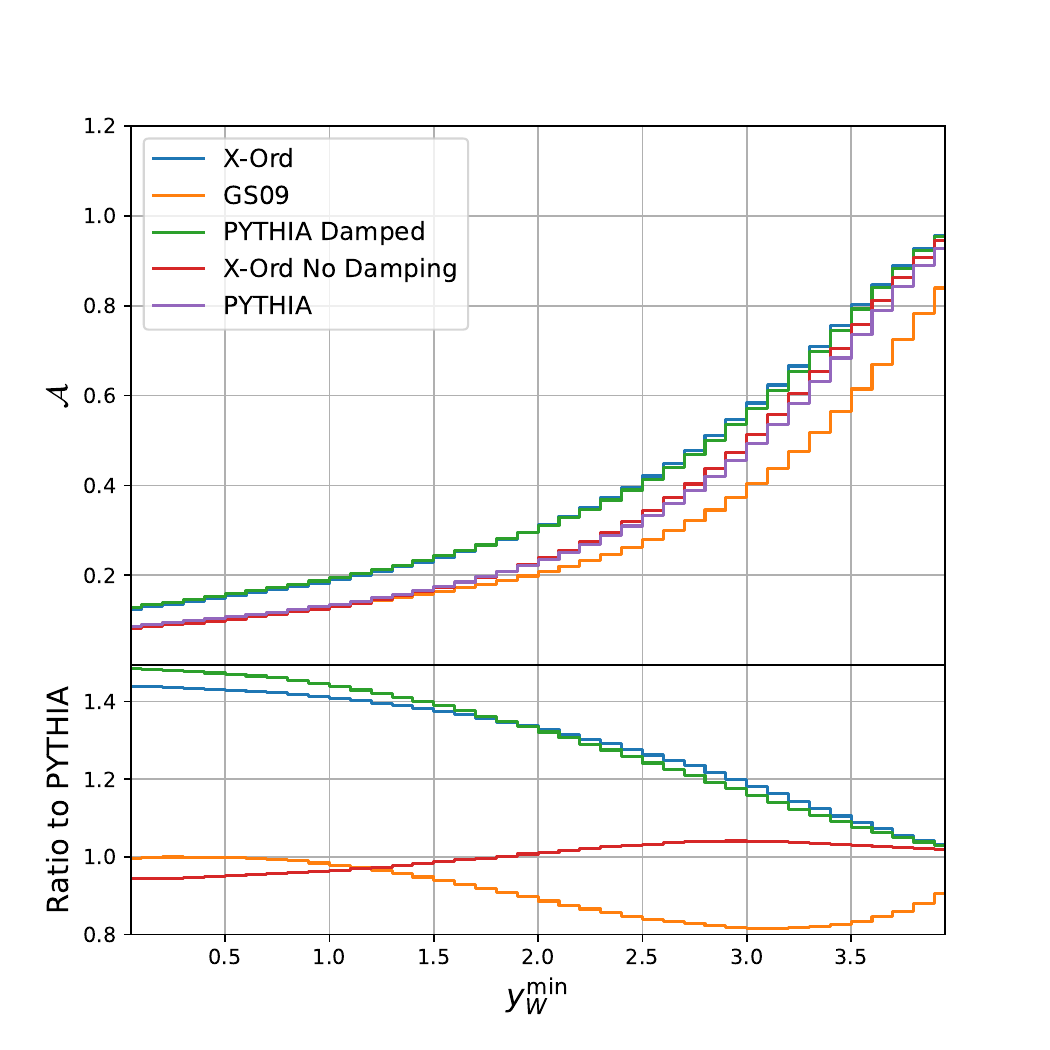}
}%
\caption{Bosonic asymmetries for same-sign $WW$ using dPDFs evaluated at $Q=2$ GeV, with the low-$x$ damping both enabled and disabled in the X-ordered and native \pythia dPDF prescriptions.}
\label{SSW:A_2GeV_damp}
\end{figure}

We traced the cause of this larger asymmetry to the substantial damping factors applied at $Q = 2$ GeV in the input dPDFs. Given the form of the damping factor, this suppresses most strongly the region where $x_1 \simeq x_2$, reducing the same-hemisphere contribution and increasing $\mathcal{A}$. In Fig.~\ref{SSW:A_2GeV_damp} we plot the bosonic asymmetries, but with the dPDFs evaluated at the ``input scale'' of $Q=2$ GeV, and we also show the impact of removing the damping from the X-ordered dPDFs, and adding the damping factors to the \pythia dPDFs. We see that this larger asymmetry for the X-ordered dPDFs is already present at the input scale, and that it is associated with the damping factors, since removing the damping factor gives a result that essentially coincides with \pythia (here again, the effect of the X-ordering procedure is rather small). The effect persists even after evolution to $Q=M_W$ (although evolution dilutes the asymmetry somewhat, particularly at small $y^{\rm min}_W$), yielding the higher asymmetries observed.

It is interesting to note that the GS09 and evolved X-ordered dPDFs give somewhat different results for the asymmetry, even though they were constructed to satisfy the same dPDF sum rules, and they are evolved with the same evolution equations to $Q=M_W$ (although one should bear in mind that they are based on different single PDF sets, MSTW 2008 LO and MMHT 2014 LO respectively). Comparing the results from the GS09 dPDFs and the (evolved) X-ordered dDPFs could be useful to get some indication of the range of possible impacts of momentum and number sum rule effects on the asymmetry (as well as other observables). 

As an aside, we investigated the impact of removing the ``$1\to 2$'' feed term on the obtained asymmetries, for the evolved X-ordered dPDFs. In same-sign $WW$ this term acts to increase the asymmetry; this is because for the $1\to2$ splitting piece, an important mechanism to generate the $qq'$ pair initiating the DPS process is a single $q$ splitting to $qg$, and then the $g$ splitting (ultimately) into $q'\bar{q}'$. If the $x$ values of both quarks in the final $qq'$ pair are large (contributing to the same hemisphere configuration), this mechanism will be somewhat suppressed, whilst if the $x$ value of the $q$ is large and the $q'$ is small (contributing to the opposite hemisphere configuration) this mechanism can be enhanced by a large chain of intermediate $g \to g$ splittings (with accompanying small $x$ logarithms) between the initially produced gluon and the final $q'$. Thus, if we remove this term, we should see a decrease in the asymmetry. This was observed in both the $W^+W^+$ and $W^-W^-$ cases, although the effect was essentially negligible here, with the asymmetries being reduced by at most $6\%$ of their initial values with the splitting included.

It is interesting to contrast the small effect of the splitting seen here with that observed in Ref.~\cite{Cabouat:2019gtm}, where a very substantial effect of removing the $1\to 2$ splitting was observed in the $W^+W^+$ process (the asymmetry nearly halves from $\sim 0.11$ to $\sim 0.06$ at $\eta_{\rm min} \simeq 0$). This difference can partially be ascribed to the account of $y$-dependence in the latter (which enhances the impact of the $1 \to 2$ feed), and the use of $n_f = 3$ rather than $n_f = 5$ in the latter (a larger $n_f$ increases the number of $qq'$ pairs with small, similar $x$ values arising from gluon-initiated $1 \to 2$ splitting processes, which in turn boosts the same-hemisphere contribution). 

However, we also investigated the effect of turning on and off the $1 \to 2$ feed contribution in the GS09 dPDFs (which have $n_f = 5$ and also do not take account of the $y$ dependence in the feed terms), and still saw a non-negligible contribution there. Turning off the feed term in $W^+W^+$ reduces the asymmetry from $0.050$ to $0.041$ at $\eta_{\rm min} \simeq 0$, and for $W^-W^-$ it reduces the asymmetry from $0.048$ to $0.041$ at $\eta_{\rm min} \simeq 0$. We found that most of the difference between the X-ordered and GS09 cases can be traced to the different starting scales for the $1 \to 2$ feed ($2$ GeV and $1$ GeV respectively) -- if one removes only the $1 \to 2$ feed between $Q=2$ GeV and $Q=M_W$ from GS09, and compares to the GS09 dPDFs with the feed, one sees a difference rather closer to the X-ordered case. This is because the $1 \to 2$ feed has a strong preference to occur ``early'' in scale (at least when neglecting the $y$ dependence) -- see Fig.~7 of \cite{Gaunt:2012dd} and the surrounding discussion. Thus, if one increases the starting scale at which the feed can contribute even only a small amount, from $1$ to $2$ GeV, this can significantly decrease its impact.

In the $W^+W^+$ case, some of the difference between the X-ordered and GS09 cases can be attributed to the different single PDFs and $\alpha_s$ values used (corresponding to the MMHT 2014 LO set rather than the MSTW 2008 LO one). The $u$ quark PDF for MSTW is somewhat larger than the MMHT one at small scales and large $x$ values, which leads to more $uq'$ pairs with asymmetric $x$ values being produced through the feed term (via the mechanism discussed above), and a larger increase of the asymmetry with the feed for GS09.

As a final small comment, we note that the asymmetries obtained from the X-ordered dPDFs are larger than those obtained with GS09 (due to the different dPDF initial conditions). Thus, if one considers the \textit{relative} impact of the $1 \to 2$ feed, one would obtain a smaller result for the X-ordered case than the GS09 one even if using the same PDFs, $\alpha_s$ and starting scale for the feed.

\subsection{Double $Z$-boson production}

\begin{figure}
    \centering
    \subfigure{%
    \label{Z:A}
    \includegraphics[width=0.5\textwidth]{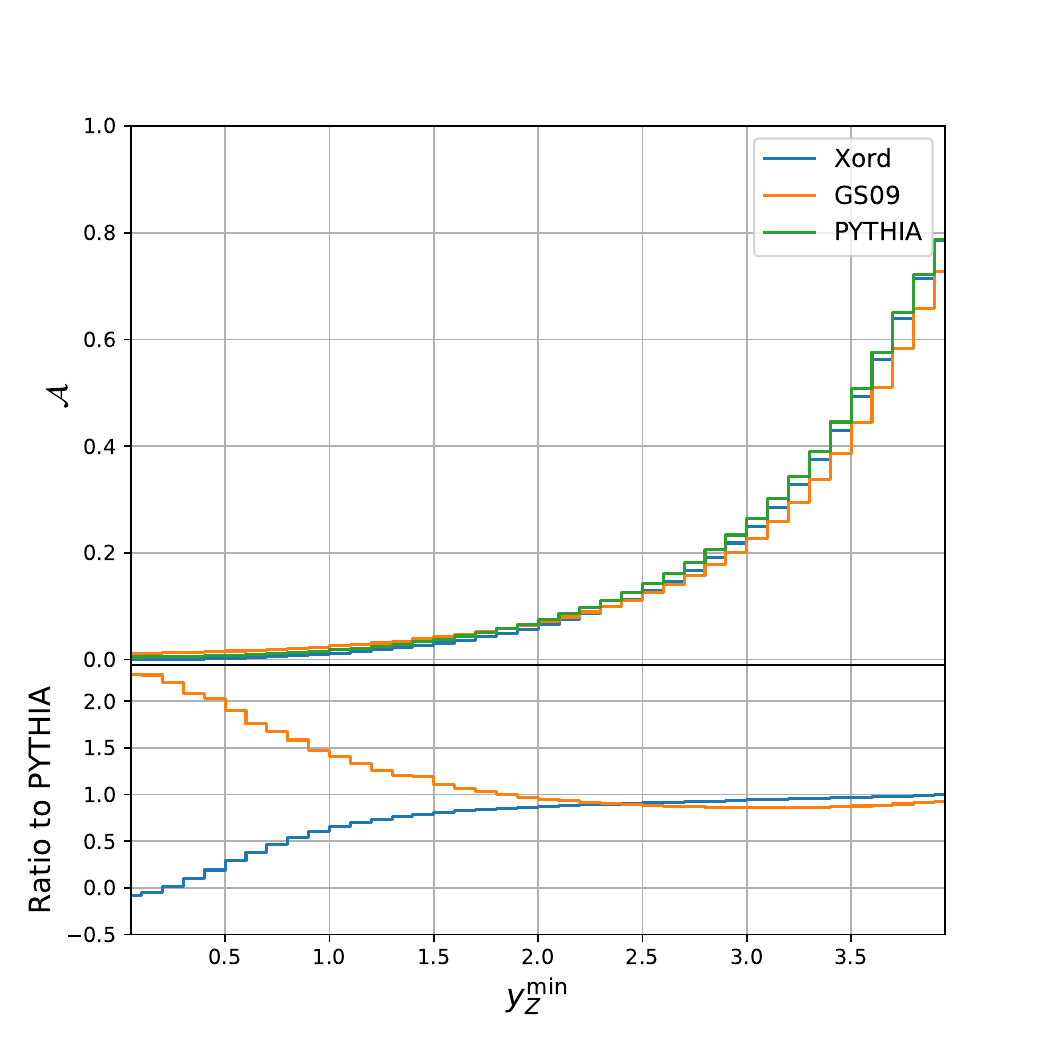}
    }
    \hspace{-9mm}
    \subfigure{%
    \label{Z:A_Zoom}
    \includegraphics[width=0.5\textwidth]{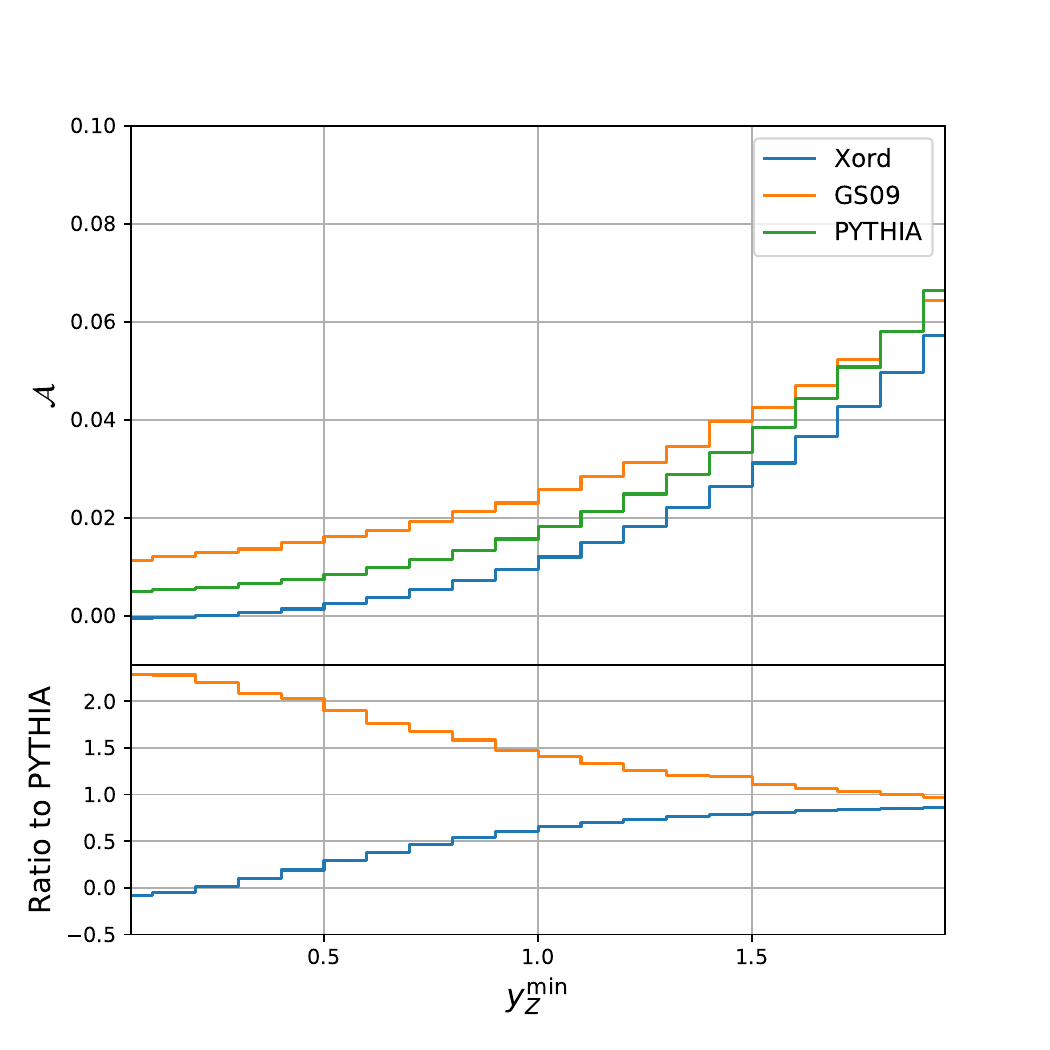}
    }%
    \caption{Bosonic asymmetries for double $Z$ boson production. The right hand plot is the same plot as the one on the left, but ``zoomed in'' to the region $0\leq y_Z^{\text{min}}\leq2$.}
    \label{Z:Asymmetries}%
\end{figure}

\begin{figure}
    \centering
    \includegraphics[width=0.8\textwidth]{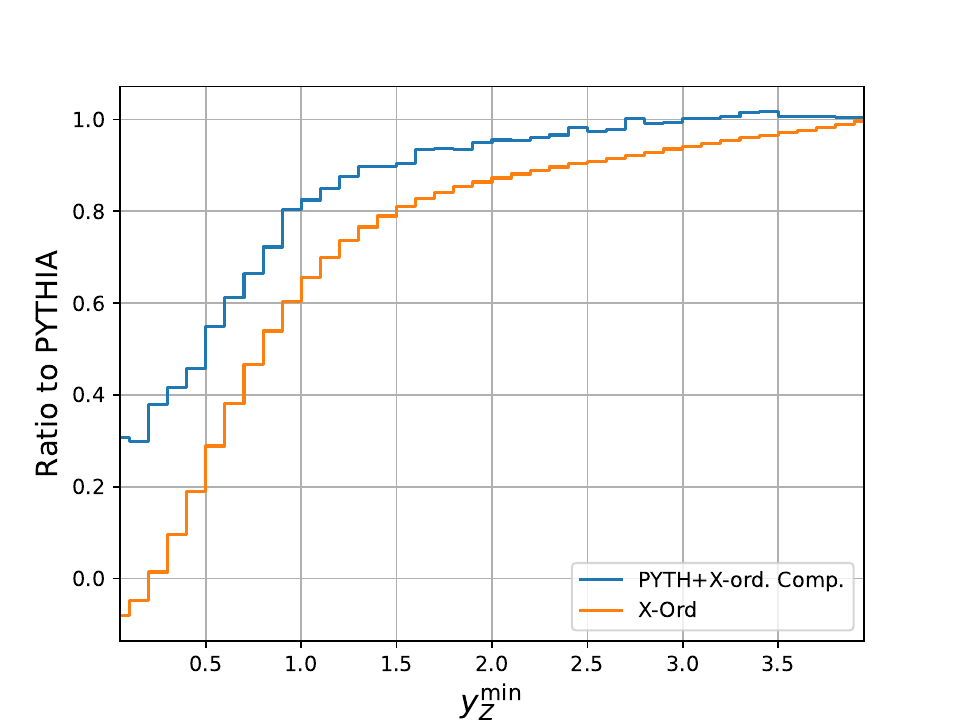}%
    \caption{Ratios of bosonic asymmetries in double $Z$ boson to \pythia. Plotted are curves for the X-ordered dPDFs, and \pythia with the default companion quark replaced by the one given in eq.~\eqref{eq:newcompanion}.}
    \label{Z:PxCompvsXord}
\end{figure}

For the production of two $Z$ bosons via DPS, we only study the bosonic asymmetry. In Fig.~\ref{Z:Asymmetries} we plot the asymmetry from the X-ordered dPDFs constructed at $Q=M_Z$, and again compare the results to \pythia and the GS09 dPDFs.
In this case we see larger differences at small $y^{\rm min}$ between the X-ordered and \pythia predictions than in the same-sign $WW$ case, although they are still not very large, decreasing $\mathcal{A}$ by $\mathcal{O}(0.01)$. This difference is largely associated with the companion quark mechanism: in Fig.~\ref{Z:PxCompvsXord} we see that when the companion mechanism in the \pythia dPDFs is swapped to the one used in the X-ordered prescription, the prediction moves a substantial way towards the X-ordered one.
The reason why the X-ordered companion mechanism drives lower asymmetries is that it has a tendency to create the companion quark closer in $x$ to the sea quark than Pythia does -- we already saw this in Fig.~\ref{ubuv_rfn_plots}. This boosts the $x_1 \simeq x_2$ region corresponding to same-hemisphere contributions, and decreases the asymmetry.

\begin{figure}
    \centering
    \subfigure{%
    \label{Z:A_2GeV}
    \includegraphics[width=0.5\textwidth]{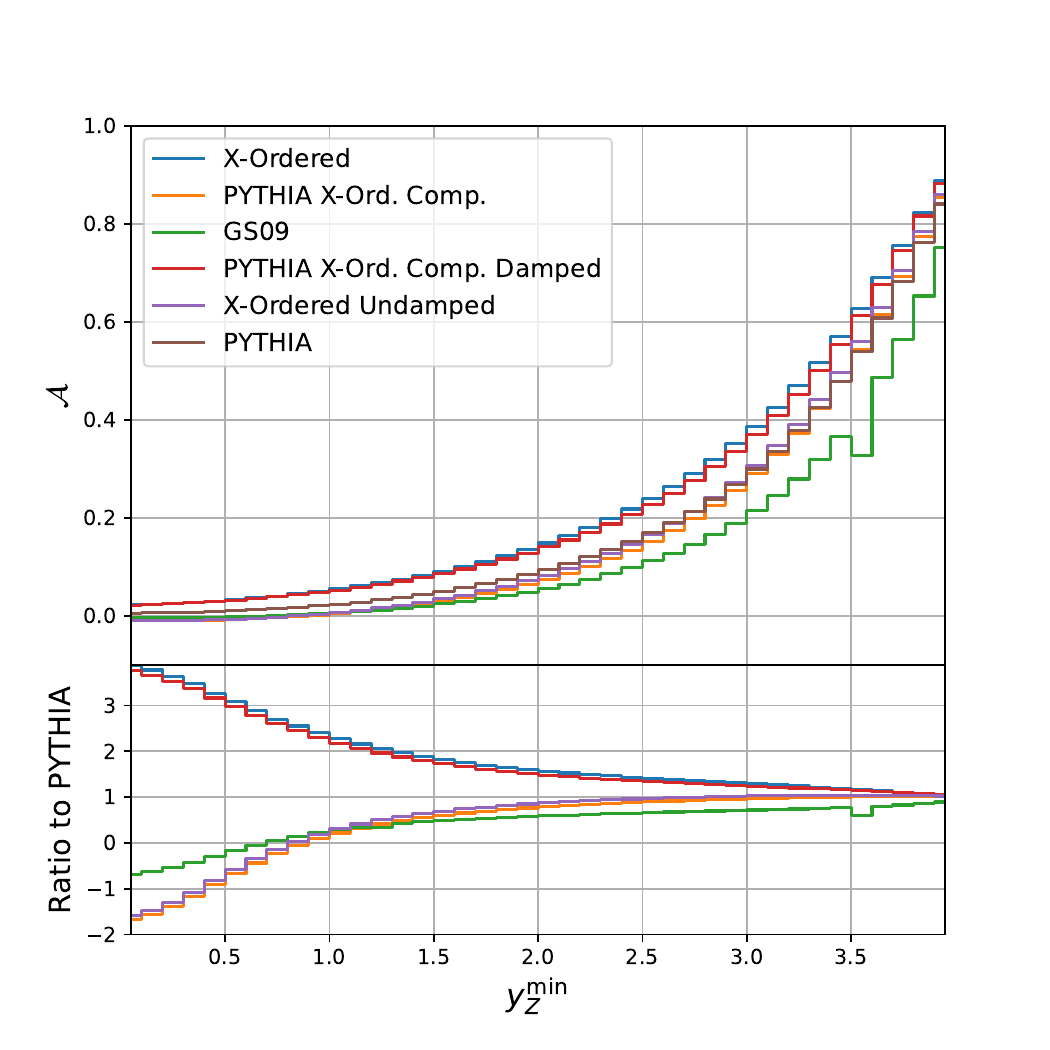}
    }
    \hspace{-9mm}
    \subfigure{%
    \label{Z:A_2GeV_Zoom}
    \includegraphics[width=0.5\textwidth]{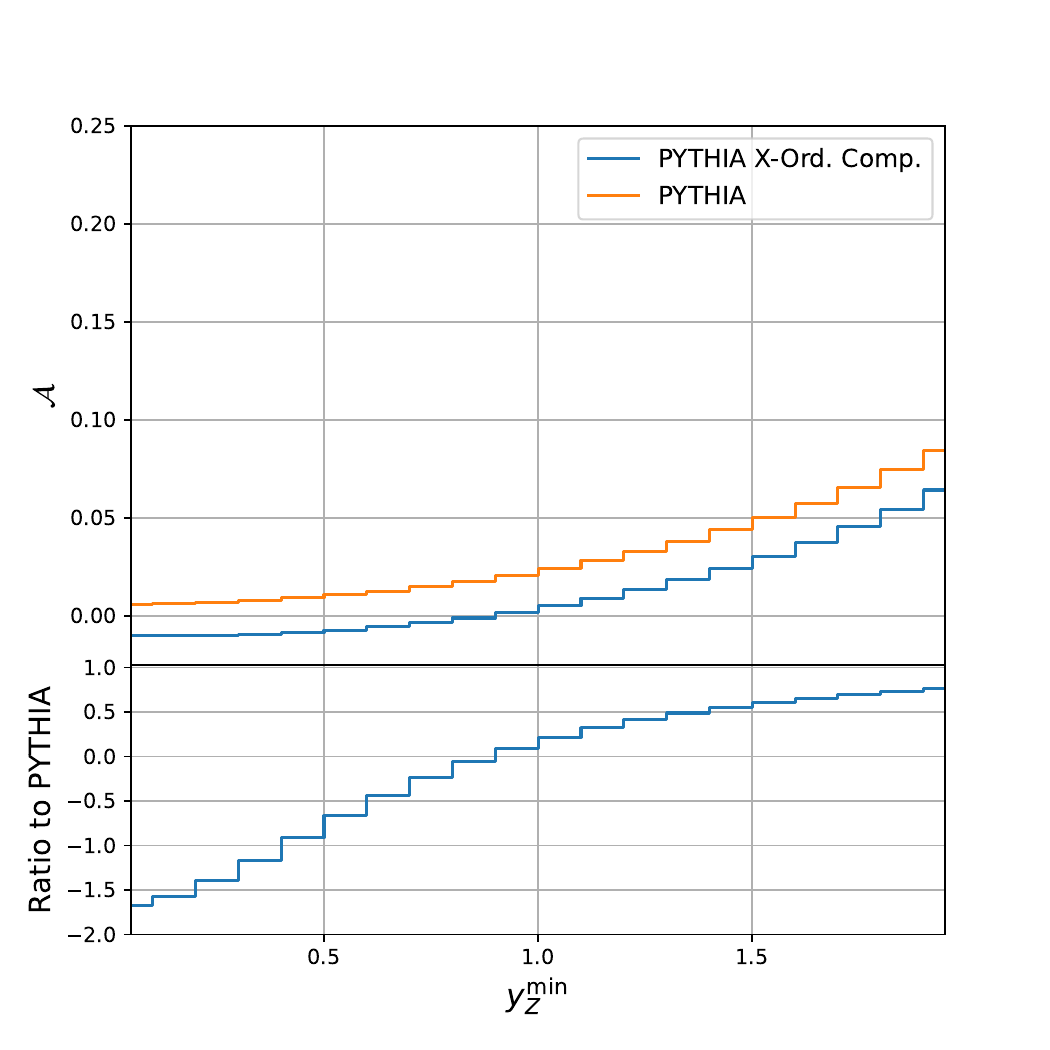}
    }%
    \caption{Bosonic asymmetries for double $Z$ production with the factorisation scale set to $2$ GeV. The left pane shows curves for the \pythia and X-ordered dPDFs, with various aspects of the X-ordered prescription enabled/disabled. The term `X-ord. comp.' means the use of the companion quark given in eq.~\eqref{eq:newcompanion}. The right pane is the same as the left, but zoomed into the region $0 \le y_Z^{\rm{min}} \le 2$, and with only two of the \pythia curves plotted.}
    \label{Z:Asymmetries_2GeV}%
\end{figure}

We will refrain from plotting the asymmetry associated with the evolved X-ordered dPDFs, since for $ZZ$ production the $1 \to 2$ splitting contributions in the evolution process play an important role (much more so than same-sign $WW$), and one should really take account of the $y$ dependence of these as well as  double counting issues with single scattering~\cite{Diehl:2017kgu}. However, we can perform the same exercise as we did for same-sign $WW$ of plotting the asymmetry with the dPDFs evaluated at the scale $Q=2$ GeV, just to give a toy illustration of how these asymmetries change when we go to lower scales. This plot is made in the left pane of Fig.~\ref{Z:Asymmetries_2GeV}, where to give additional insight into how the different aspects of the X-ordering prescription affect the asymmetry, we also plot curves for Pythia with various aspects of the X-ordering prescription included, and the X-ordered dPDFs with the damping factor removed. We observe that the modification producing the most significant effect on the asymmetry is the damping factor, where inclusion of the damping factor increases the asymmetry considerably. This is very similar to what we saw for the same-sign $WW$ case at $Q = 2$ GeV. The change in the companion mechanism has a slightly larger impact at $Q = 2$ GeV than it does at $Q = M_Z$, shifting the asymmetries downwards at lower $y_Z^{min}$ by $\sim 0.02$ (see the right pane of Fig.~\ref{Z:Asymmetries_2GeV}). 

\section{Summary and Outlook}
\label{sec:summary}
The equal scale $m$-parton distribution functions (mPDFs) express the probability to find $m$ partons in the proton with given flavours $j_i$ and momentum fractions, $x_i$, when all partons are probed at the same scale $Q$. These distributions must satisfy two simple requirements: first, they must be symmetric when swapping over the flavours and momentum fractions of any two partons, $\{x_i,j_i\} \leftrightarrow \{x_k,j_k\}$. Second, they must satisfy number and momentum sum rules, which relate the integrals of the mPDFs over one of the momentum fraction arguments to mPDFs with one less parton. 
The \pythia procedure for constructing model mPDFs out of single PDFs, introduced by Sj\"ostrand and Skands \cite{Sjostrand:2004pf, Sjostrand:2004ef}, is based on an ``ordering'' of the interactions in energy, and cannot straightforwardly be used to construct equal scale mPDFs satisfying both of these constraints simultaneously. One can either obtain non-symmetric mPDFs perfectly satisfying the sum rules in only one parton argument, or, naively symmetrising these mPDFs, one obtains symmetric mPDFs that satisfy the sum rules imperfectly. The extent to which the sum rules are violated can be quite significant ($\mathcal{O}(100\%)$ or more), particularly when the $x$ fractions of the partons that are not integrated over are large \cite{Fedkevych:2022myf}.

In this paper we have developed an algorithm for constructing equal scale mPDFs that has as its starting point the \pythia procedure, but incorporates a number of modifications which enforce the symmetry of the mPDF, and are intended to improve the extent to which the mPDFs satisfy the sum rules vs the naively symmetrised \pythia mPDFs. We introduced three modifications: first, we altered the \pythia ``companion quark'' mechanism associated with $g \to q\bar{q}$ splitting processes such that it is naturally symmetric and does not double count the gluon splitting process (essentially following Ref.~\cite{Gaunt:2009re}). Second, we imposed an ``$x$-ordering'' for the partons in the \pythia procedure, with the highest $x$ parton being picked first, and so on. A smoothing procedure is applied when the $x$ values of two (or more) partons approach each other, to avoid discontinuities in the mPDF. The rationale behind this ``$x$-ordering'' procedure is the intuitive expectation that small $x$ partons should be affected more by the presence of large $x$ partons than vice versa. Finally, we introduced a damping function, that suppresses the mPDF when there are two or more sea quarks or gluons in the mPDF with small $x$ values (and the suppression increasing the more small $x$ partons there are in the mPDF). In order to best satisfy the sum rules at a range of different scales, the damping is scale-dependent, with stronger damping at smaller $Q$. We refer to the mPDFs resulting from this procedure as the ``X-ordered mPDFs''.

We tested the extent to which the sum rules are satisfied by this algorithm for the explicit cases of $m=2$, the double PDF or dPDF, and $m=3$, the triple PDF or tPDF. In this investigation the single PDFs used in the construction were the MMHT 2014 LO set. In the dPDF case we tested the X-ordered construction at a high scale $Q=M_Z$ and at a low scale $Q=2$ GeV.  At $Q=M_Z$ all sum rules are satisfied to well within $10\%$ over a broad range of $x$ values for the non-integrated parton, $10^{-6} < x_1 \le 1$. This represents an improvement over the naive symmetrised \pythia dPDFs, which violate the sum rules at a level much greater than $10\%$ at large $x_1$ \cite{Fedkevych:2022myf}. At the lower scale $Q=2$ GeV the extent to which the sum rules are satisfied is slightly worse, and for some sum rules there is a small region of phase space where the sum rules are violated by slightly more than $10\%$, although these violations are in practice not so relevant (since they are associated with a sea parton having large $x_1 \sim 0.1$). We also considered the use of the low scale dPDFs at $2$ GeV as an input to pQCD evolution, evolving them to $Q = M_Z$ according to the inhomogeneous double DGLAP equation, and studying how well the evolved dPDFs satisfied the sum rules. We found that evolution smoothed the sum rule curves and brought all sum rules to within $10\%$.

In the tPDF case we restricted our tests to the tPDFs constructed at $Q=M_Z$. We found that all momentum sum rules are obeyed to within $10\%$, and that the number sum rules are obeyed to within $10\%$ over most of the phase space for the two non-integrated partons, with any violations only being mildly above this level. The only exception to this is the $guu_v$ number sum rule, where the number sum rule violations are slightly worse than $20 \%$ when the $x$ values of the two non-integrated partons are very small. Again, this represents a notable improvement with respect to the naively symmetrised \pythia tPDFs, where there are much more significant violations of the sum rules.

We performed a simple toy study of the extent to which the modifications in the X-ordered dPDFs impact predictions at the level of double parton scattering (DPS) cross sections, comparing also against predictions from the GS09 dPDFs \cite{Gaunt:2009re}. We chose to study same-sign $W^\pm W^\pm$ and $ZZ$ production, and restricted our attention to the boson/lepton rapidity asymmetry observables, since these are sensitive to correlations in the dPDFs. It was found that there were two main modifications producing differences in the asymmetries from the \pythia predictions: the damping, if it is applied in the construction of the dPDFs at a low scale before the dPDF is evolved up to $M_W/M_Z$ (since in this case the damping is strongest), and the modification to the companion quark mechanism. The (low-scale) damping causes a substantial increase in the asymmetry since it suppresses the same-hemisphere $x_1\sim x_2$ region, whilst the change in the companion quark causes a smaller decrease in the asymmetry, since the new companion has a stronger preference to be close in $x$ to its sea partner, and boosts the $x_1 \sim x_2$ region. It is interesting to note the substantial difference in results between the GS09 dPDFs and the evolved X-ordered dPDFs, even though these dPDFs are constructed to satisfy the same sum rules and have been evolved using the same pQCD evolution equations - by comparing these dPDFs one can get a rough indication of the different possible impacts of number and momentum sum rule effects on DPS distributions.

There are various ways in which the results in this paper can be used in further work. The constructed X-ordered mPDFs could be directly used in approximate multiple scattering cross section formulae (like Eq.~\eqref{eq:dPDF_Scatter_noy}); this approach neglects correlations in the transverse space $y$ between the partons (as well as potential single/double/triple scattering double counting issues) and doesn't properly account for pQCD evolution effects, but it is a simplest approach to account for number and momentum sum rule effects. It is worth noting that our X-ordered tPDFs represent the first set that has been constructed to be fully symmetric in the parton arguments and to satisfy the sum rules to a good level of precision over the full kinematic range. This is pertinent given the increasing activity with regards to predicting and measuring TPS cross sections at the LHC (see \textit{e.g.} Refs.~\cite{dEnterria:2017yhd, Shao:2019qob, CMS:2021qsn, Maneyro:2024twb}).

Alternatively, our mPDFs could be used as a low-scale input in pQCD evolution to larger scales (as we have explored to some extent ourselves in the above). The optimal approach would be to extend our model mPDFs to include $y$-dependence (possibly following Ref.~\cite{Diehl:2020xyg}), and then evolve these to higher scales using the full $y$-dependent framework, using for example {\tt ChiliPDF} \cite{Diehl:2023cth} (for the case of the double parton density). The resulting $y$-dependent double parton densities could then be used in the full DPS cross section formula 
Eq.~\eqref{eq:dPDF_Scatter}, or in the DPS parton shower simulation dShower \cite{Cabouat:2019gtm}.

A final future direction would be to explore the extent to which the modifications described here could be incorporated into the \pythia MC. This would not be straightforward since we have only considered the equal-scale mPDF case here, whilst the \pythia model has to cover the general unequal scale case. One would need to consider the appropriate sum rules in the unequal scale case, generalize the mPDF construction to this case, and phrase the construction in a way that is compatible with the \pythia energy-ordered MPI model, which may not be readily possible (outside some kind of reweighting-based approach, which will also presumably become impractical for large numbers of scatters). One modification that could presumably be incorporated without too much work is the implementation of the companion quark as in Eq.~\ref{eq:newcompanion}. It could be interesting to include this as an option in \pythia, and investigate the impact on observables sensitive to multiple parton scattering.